\documentclass[twocolumn,times,twocolappendix]{aastex701}

\begin{document}

\title{Complex Nuclear  Structure in Seyfert 2 Galaxy NGC 4388 Revealed by XRISM Observation}

\author[0009-0006-4377-4219]{Kanta Fujiwara}
\affiliation{Department of Astronomy, Kyoto University, Kitashirakawa-Oiwake-cho, Sakyo-ku, Kyoto, 606-8502, Japan}
\email{fujiwara@kusastro.kyoto-u.ac.jp}

\author[0000-0002-5485-2722]{Yoshihiro Ueda}
\affiliation{Department of Astronomy, Kyoto University, Kitashirakawa-Oiwake-cho, Sakyo-ku, Kyoto, 606-8502, Japan}
\email{ueda@kusastro.kyoto-u.ac.jp}

\author[0000-0002-5701-0811]{Shoji Ogawa}
\affiliation{Institute of Space and Astronautical Science (ISAS), Japan Aerospace Exploration Agency (JAXA), 3-1-1 Yoshinodai, Chuo-ku, Sagamihara, Kanagawa 252-5210}
\email{ogawa@kusastro.kyoto-u.ac.jp}

\author[0009-0000-9577-8701]{Yuya Nakatani}
\affiliation{Department of Astronomy, Kyoto University, Kitashirakawa-Oiwake-cho, Sakyo-ku, Kyoto, 606-8502, Japan}
\email{nakatani@kusastro.kyoto-u.ac.jp}

\author[0000-0003-2869-7682]{Jon M. Miller}
\affiliation{Department of Astronomy, University of Michigan, MI 48109, USA}
\email{jonmm@umich.edu}

\author[0000-0002-6054-3432]{Takashi Okajima}
\affiliation{NASA Goddard Space Flight Center (GSFC), Greenbelt, MD 20771, USA}
\email{takashi.okajima@nasa.gov}

\author[0000-0002-6808-2052]{Taiki Kawamuro}
\affiliation{Department of Earth and Space Science, Osaka University, 1-1 Machikaneyama, Toyonaka 560-0043, Osaka, Japan}
\email{kawamuro@ess.sci.osaka-u.ac.jp}

\author[0000-0001-9379-4716]{Peter G Boorman}
\affiliation{Max-Planck-Institut für Extraterrestrische Physik, Gießenbachstraße 1, 85748 Garching, Germany}
\email{boorman@mpe.mpg.de}

\author[0009-0006-4968-7108]{Luigi Gallo}
\affiliation{Department of Astronomy and Physics, Saint Mary’s University, Nova Scotia B3H 3C3, Canada}
\email{Luigi.Gallo@smu.ca}

\author[0000-0003-2161-0361]{Misaki Mizumoto}
\affiliation{Science Research Education Unit, University of Teacher Education Fukuoka, 1-1 Akama-bunkyo-machi, Munakata, Fukuoka
811-4192, Japan}
\email{mizumoto-m@fukuoka-edu.ac.jp}

\author[0000-0002-7962-5446]{Richard Mushotzky}
\affiliation{Department of Astronomy University of Maryland College Park Md 20742}
\email{rmushotz@umd.edu}

\author[0000-0001-6020-517X]{Hirofumi Noda}
\affiliation{Astronomical Institute, Tohoku University, 6-3 Aramakiazaaoba, Aoba-ku, Sendai, Miyagi 980-8578, Japan}
\email{hirofumi.noda@astr.tohoku.ac.jp}

\author[0000-0003-1780-5481]{Yuichi Terashima}
\affiliation{Ehime University, Graduate School of Science and Engineering, 2-5, Bunkyo-cho, Matsuyama-shi, Ehime,
790-8577, Japan}
\email{terashima.yuichi.mc@ehime-u.ac.jp}

\author[0000-0002-6562-8654]{Francesco Tombesi}
\affiliation{INAF – Astronomical Observatory of Rome, Via Frascati 33, 00040 Monte Porzio Catone, Italy}
\affiliation{INFN - Rome Tor Vergata, Via della Ricerca Scientifica 1, 00133 Rome, Italy}
\email{francesco.tombesi@roma2.infn.it}

\author[0000-0002-5488-1961]{Bert Vander Meulen}
\affiliation{European Space Agency (ESA), European Space Research and Technology Centre (ESTEC), Keplerlaan 1, 2201 AZ Noordwijk, The Netherlands}
\email{Bert.VanderMeulen@esa.int}

\author[0000-0002-9754-3081]{Satoshi Yamada}
\affiliation{The Frontier Research Institute for Interdisciplinary Sciences, Tohoku University, Aramaki, Aoba-ku, Sendai, Miyagi 980-8578, Japan}
\affiliation{Astronomical Institute, Tohoku University, Aramaki, Aoba-ku, Sendai, Miyagi 980-8578, Japan}
\email{satoshi.yamada@terra.astr.tohoku.ac.jp}

\begin{abstract}

We report results from the simultaneous XRISM (183 ks) and NuSTAR (62 ks) observations of the Seyfert-2 galaxy NGC 4388.  This AGN has the
brightest Fe K$\alpha$ line among Compton-thin, obscured sources.  To
model the reflection continuum and fluorescent lines, we employ an
updated version of XCLUMPY 
and a broad line region model with a disk-like geometry .
The profile of the neutral Fe-K fluorescent line is well
described as the sum of three components convolved with Gaussians with
FWHM values of $\sim 290\ \mathrm{km\ s^{-1}}$, $\sim
1470\ \mathrm{km\ s^{-1}}$, and $\sim
11100\ \mathrm{km\ s^{-1}}$.  These line widths correspond to radii of
1.5 pc, 0.060 pc, and $1.0\times10^{-3}$ pc by assuming Keplerian motion, which we
interpret as the dusty torus, its inner edge region, and the BLR,
respectively.  The data suggest that the Fe K$\alpha$ BLR component is
larger than that of H$\alpha$ (FWHM of 4500 $\mathrm{km\ s^{-1}}$) in the polarized optical
spectrum, implying that the velocity field of the BLR is dominated by
that parallel to the equatorial plane. In addition, Fe XXVI Ly$\alpha$
and Fe XXV absorption lines are detected, characterized by $\log{\xi}
\sim 3.50~\mathrm{erg\ cm\ s^{-1}}$, $\log{N_{\mathrm{H}}} \sim
22.1~\mathrm{cm^{-2}}$, $v_{\mathrm{out}} \sim
40\ \mathrm{km\ s^{-1}}$, and $\sigma_v \sim
160\ \mathrm{km\ s^{-1}}$. We infer that the absorber is
gravitationally bound and is possibly associated with a failed wind,
consistent with a radiation-driven fountain flow.

\end{abstract}

\keywords{Active galactic nuclei (16), Astrophysical black holes (98), High energy astrophysics (739), Seyfert
 galaxies (1447), Supermassive black holes (1663), X-ray active galactic nuclei (2035)}


\section{Introduction} 

Understanding the structure of active galactic nuclei (AGN) is
essential for revealing the physical mechanisms that govern the growth
of supermassive black holes (SMBH).  The distribution and kinematics
of circumnuclear gas regulate accretion onto the SMBH and mediate
energy and momentum feedback to the host galaxy.  Constraining the
geometry and dynamics of this multi-phase medium is therefore a
central goal of modern AGN studies.

In type 2 AGN, direct emission from the central engine is obscured by
a geometrically thick absorber, commonly interpreted as a ``dusty'' torus (e.g., \citealt{2017NatAs...1..679R}).  
This torus obscures  inner structure such as the broad line
region (BLR), making investigations of the innermost nuclear
environment historically reliant on polarimetric observations
\citep{1985ApJ...297..621A}. 
 X-rays, with their strong penetrating
power, can probe material located inside the torus, including the BLR
and accretion disk (\citealt{1989MNRAS.238..729F, 1995MNRAS.277L..11F,
  2006ApJ...652.1028B}), as far as the absorption is Compton thin
(i.e., a line-of-sight absorption is $N_{\rm H} \lesssim 1.5 \times
10^{24}$ cm$^{-2}$).
Moreover, we also study the torus itself through the
line-of-sight absorption of the intrinsic X-ray emission and reflected
X-rays accompanied by prominent narrow fluorescent lines (e.g.,
\citealt{2016ApJ...831...37K, 2018ApJ...853..146T}). 

In practice, however, the limited energy
resolution of CCD detectors has made it difficult to separate multiple
velocity components in the Fe-K band to isolate contributions from
the torus, BLR, and the accretion disk.

The microcalorimeter onboard X-Ray Imaging and Spectroscopic Mission;
XRISM (\citealt{2021SPIE11444E..22T, 2025PASJ..tmp...28T}), Resolve 
(\citealt{ishisaki2025, kelley2025})
overcomes this limitation by providing $\sim$ 5 eV
resolution across the Fe-K band, sufficient to resolve line widths
corresponding to a few hundred km s$^{-1}$. 
This enabled the detection of BLR-associated Fe K$\alpha$ emission not only
in type 1–1.5 AGN such as NGC 4151 \citep{2024ApJ...973L..25X}, Mrk
279 \citep{2025ApJ...994L..10M}, NGC 7213 \citep{2025ApJ...994L..13K}
and NGC 3783 \citep{2026A&A...706A.255L} but also in the type 2 AGN
Centaurus A (\citealt{2025PASJ...77S.209B}; Nakatani et al. in prep.), demonstrating that Resolve can isolate inner nuclear
components even when the direct continuum is obscured.
Thus, XRISM/Resolve enables simultaneous and self-consistent
investigation of both the outer circumnuclear structure—dominated by
torus reflection—and the inner regions, including BLR-scale
fluorescence and highly ionized winds, even in type 2 AGN.  This
represents a major step forward in mapping the full radial structure
of AGN using physically grounded spectroscopy.

NGC~4388 is a nearby Seyfert-2 galaxy at a distance of $19\pm2$~Mpc
\citep{1982AJ.....87.1628H}.  Its X-ray spectrum is characteristic of
a Compton-thin type-2 AGN, absorption with a line-of-sight column density of $\log
N_{\rm H}/{\rm cm^{-2}} \simeq 23.8$ (e.g.,
\citealt{2016ApJS..225...14K}).  
A prominent Fe~K$\alpha$ fluorescence line was first reported from the ASCA observation by \citet{1997MNRAS.285..683I}.  Among Compton-thin Seyfert~2
galaxies, NGC~4388 exhibits the brightest narrow Fe~K$\alpha$ line,
with a flux of $8.3\times10^{-5}\ {\rm photons\ cm^{-2}\ s^{-1}}$
(\citealt{2011ApJ...727...19F}).  Because the absorption is moderate
rather than Compton-thick, emission from the BLR
can be observed directly in X-rays.  In addition,
relativistically broadened Fe K$\alpha$ emission from the
accretion disk has been investigated, although its presence
remains uncertain (\citealt{2016ApJS..225...14K, 2017ApJ...843...89K,
  2023MNRAS.522..394Y}).

NGC~4388 is also a known H$_2$O megamaser source, indicating that the
accretion disk is viewed at a nearly edge-on inclination, a geometry
further supported by the observed jet orientation
(\citealt{2011ApJ...727...20K}).  Such a high inclination maximizes
Doppler broadening of fluorescent emission from the torus, enhancing
the detectability of kinematic structure in the Fe-K band.  The
combination of an extremely bright narrow Fe~K$\alpha$ line,
Compton-thin absorption, and high inclination makes NGC~4388 one of
the most favorable targets for probing the torus and inner nuclear
regions through high-resolution X-ray spectroscopy.

The black hole mass has been estimated to be
$8.5\times10^{6}\ M_{\odot}$ from megamaser dynamics
(\citealt{2011ApJ...727...20K}).  Together with the time-averaged
Swift/BAT luminosity ($L_{2-10~\mathrm{keV}} = 10^{43.1}$ erg cm s$^{-1}$; \citealt{2017ApJS..233...17R}), this implies an Eddington ratio of $\sim0.1$,
typical of local Seyferts.  Thus, the physical conditions in NGC~4388 may be considered broadly
representative of the majority of nearby AGN.
Moreover, NGC 4388 is one of the very few type 2 AGN in which a warm
absorber has been reported  around 6--7 keV, based on NICER and Chandra observations
(\citealt{2019ApJ...884..106M, 2024ApJ...966...57G}).
 They detected a highly photoionized wind ($\log{\xi}$/(erg cm s$^{-1}$) $\sim$ 3.4), but its outflow velocity could not be well constrained due to the limited energy resolution of NICER and Chandra.
These features  making it a rare
and valuable laboratory for exploring the connection between the
torus, BLR, and ionized outflows.

This paper presents the analysis of the first XRISM observation of NGC~4388.
 The goal of this work is to unveil the nuclear structure of a type-2 AGN, including the torus, ionized wind, and BLR, through high-resolution X-ray spectral analysis. 
We describe the data reduction procedures for the XRISM and NuSTAR observations in Section~\ref{sec:reduction}.  The
methodology of our physically motivated spectral modeling and
simulations is presented in Section~\ref{sec:modeling}.
Section~\ref{sec:fitting} details the spectral analysis techniques and the results.
Their implications for the structure of the
circumnuclear region in NGC~4388 are discussed in
Section~\ref{sec:discussion}.
Throughout this paper, errors correspond to the 90$\%$
confidence region for a single parameter.

\section{Observations and Data Reduction}
\label{sec:reduction}

\begin{figure}[htb]
\centering
\includegraphics[width=0.45\textwidth]{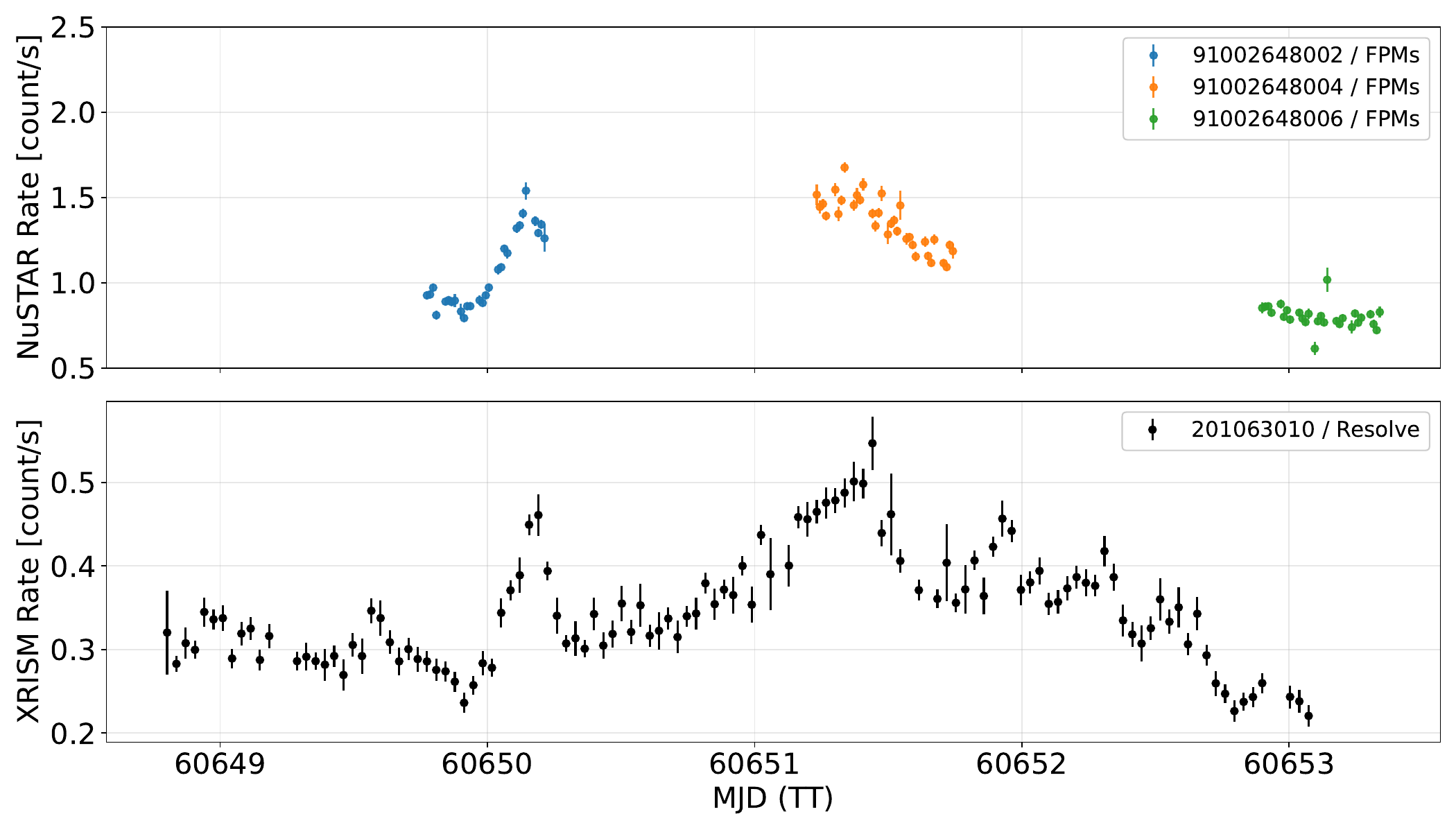}
\caption{
The light curves obtained with XRISM/Resolve and NuSTAR/FPMs.
The NuSTAR light curve (top) is extracted in the 8–70  keV band, while the XRISM/Resolve light curve (bottom) is extracted in the 3–10 keV band.
\label{figure:light-curve}}
\end{figure}

XRISM observed NGC 4388 as a GO1 target (PI: Ueda) from
2024 December 4 to December 9 (ObsID 201063010), obtaining a net exposure of 183 ks.  During the same period, NuSTAR performed three
coordinated observations (ObsIDs 91002648002/004/006) with individual
exposures of 21, 22, and 19 ks, respectively, yielding a combined
exposure of 62 ks. The observation log is listed in table \ref{tab-obj}.
The light curves obtained with XRISM/Resolve and NuSTAR/FPMs are shown in Figure~\ref{figure:light-curve}.
We focus on time-averaged spectroscopy to achieve the highest S/N in the high-resolution XRISM spectrum, 
even when the count rate is variable.
We also confirmed that the broadband continuum and fine spectral features, such as absorption lines, do not show significant variability.

\subsection{XRISM}

XRISM carries two focal-plane instruments: the
X-ray microcalorimeter \textit{Resolve}
(\citealt{ishisaki2025, kelley2025})and
the X-ray CCD camera \textit{Xtend} \citep{2025PASJ...77S..10N}, which
are coupled to the X-ray Mirror Assemblies (\textit{XMA}). In this work,
we analyze only the Resolve data to highlight the new spectroscopic
capabilities of XRISM.

The Resolve data were processed following the standard procedures
described in the \textit{XRISM ABC Guide} (v1.0), using
\textsc{HEAsoft} v6.35.1 and the XRISM CALDB (released in March 2025).  We
first applied the recommended event screening using the
\texttt{RISE\_TIME}, \texttt{DERIV\_MAX}, \texttt{ITYPE}, and
\texttt{STATUS} flags, and created a cleaned Level~2 event file with
the \texttt{ftcopy} command.
We then used \textsc{Xselect} to extract the source events. Following
the ABC Guide, we selected only the high spectral--resolution events
by applying \texttt{GRADE="0:0"}, i.e., using Hp events only.  In
addition, we excluded detector pixel 27, which are known to
have calibration uncertainties \citep{eckart2025}, and retained all other pixels for the spectral extraction.
We also excluded detector pixel 12, because it is a calibration pixel.  
Although Resolve nominally covers the 0.3--12~keV band, the low-energy
range ($<3$~keV) is strongly suppressed because the gate valve remained
closed during the observation, resulting in insufficient photon
statistics.  Therefore, we restricted the spectral analysis to the
3--10~keV band, where the signal-to-noise ratio is sufficiently high
and the calibration is robust.

We generated the high-resolution response matrix file (RMF) using
\texttt{rslmkrmf} and created the exposure map with \texttt{xaexpmap}.
We then produced the ancillary response file (ARF) using
\texttt{xaarfgen}, which performs a ray-tracing simulation including
mirror reflection, transmission, and detector effects.  The simulation
assumed a point source located at (RA = 186.4450$^\circ$, Dec =
12.6620$^\circ$) and used the same detector-coordinate region as adopted
for the spectral extraction.  The number of simulated photons was set
to 300{,}000, and all calibration components (QE, contamination, gate
valve, mirror model, scattering, obstruction, etc.) were taken from
CALDB. Given the intrinsically low non--X-ray background (NXB) of
Resolve, we did not apply an explicit NXB subtraction.  As a
consistency check, we verified that no known NXB lines appear in the
extracted spectrum.

\subsection{NuSTAR}

NuSTAR has two detectors, FPMA and FPMB, which cover the 3–79 keV energy range. The FPM data were processed with HEAsoft v6.35.1 and the NuSTAR CALDB released on 2021 November. We extracted the source spectra from a circular region with a 2$'$ radius centered on the source peak. The background was taken from a nearby source-free circular region of the same radius. The source spectra, background spectra, and the RMF and ARF files from the two modules were combined using \texttt{ADDASCASPEC}.
Although the two detectors have independent calibrations and are often analyzed separately, we confirmed that fitting them individually provides consistent results. 

\begin{deluxetable*}{lcccccc}
\tablewidth{\textwidth}
\tablecaption{Observation log \label{tab-obj}}
\tablehead{
(1) & (2) & (3) & (4) & 
}
\startdata
XRISM & 201063010 & 2024 Dec 04 18:50  & 183 \\
NuSTAR & 91002648002 & 2024 Dec 05 18:26 & 21 \\
    & 91002648004 & 2024 Dec 07 05:36  & 22 \\	
    & 91002648006 & 2024 Dec 08 21:31 & 19 \\
\enddata
\tablecomments{
(1): observatory. (2): observation identification number. (3): start date and time. 
(4): exposure time in units of kiloseconds.}
\end{deluxetable*}

\section{Model Components}
\label{sec:modeling}

\subsection{Torus}
\label{sec:torus}

\begin{figure}[htb]
\centering
\includegraphics[width=0.45\textwidth]{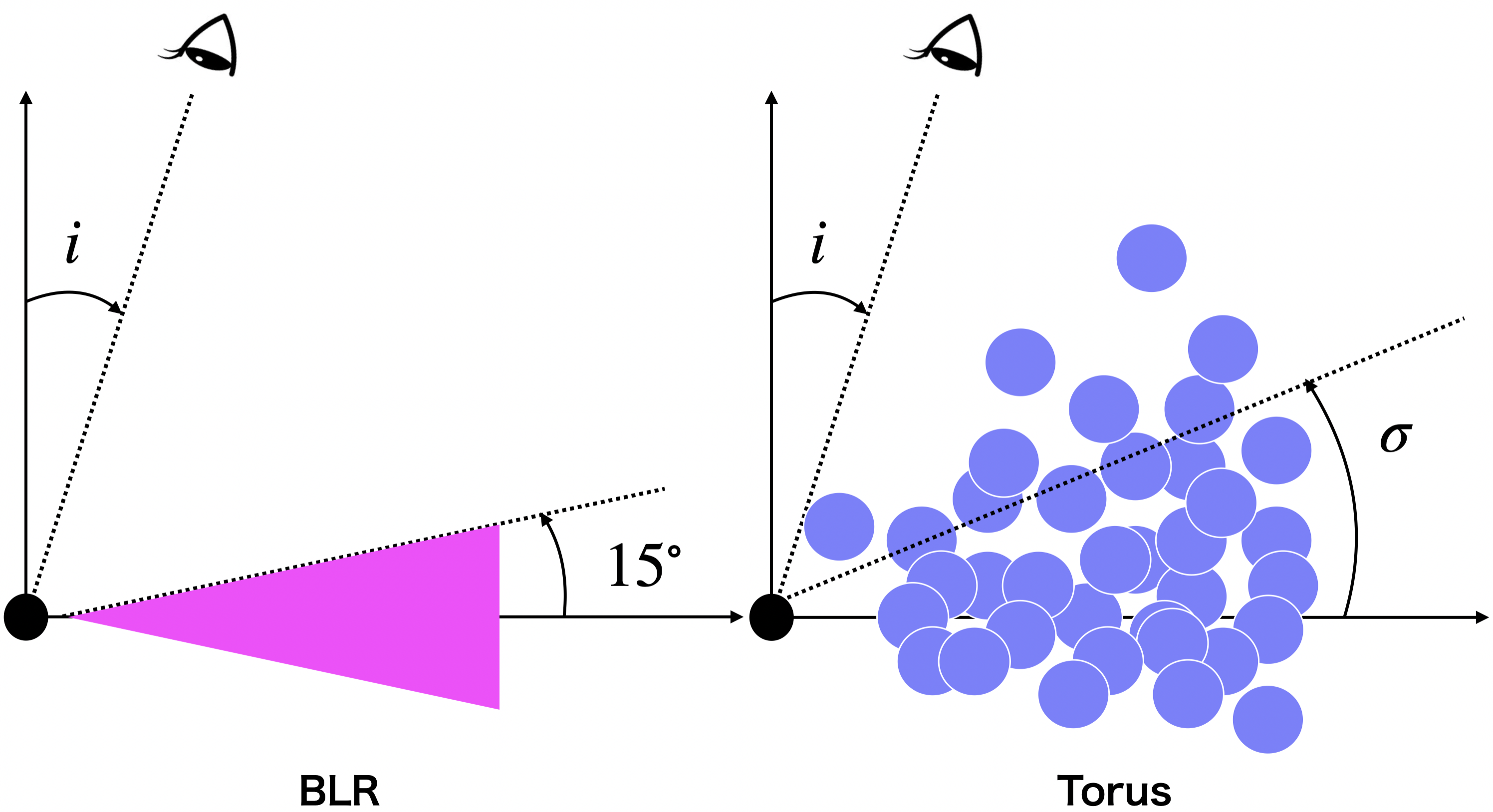}
\caption{
Geometry assumed in our reflection modeling.
The BLR (left) is treated as a uniform-density wedge with a fixed opening angle, while the torus (right) follows a clumpy structure consistent with the XCLUMPY model.
\label{figure:model_geomety}}
\end{figure}

The XCLUMPY model (\citealt{2019ApJ...877...95T}) is an 
X-ray reflection  model that assumes a clumpy torus geometry (Figure
\ref{figure:model_geomety} Right), 
which has three free parameters: torus angular width, inclination, and hydrogen column density along a line-of-sight on the equatorial plane. 
The torus is composed of randomly placed
clumps according to a power-law distribution in the radial direction and a Gaussian distribution in the polar direction. The number
density function $d(r, \theta, \phi)$ is represented in the spherical coordinate system (where $r$ is radius, $\theta$ is polar angle, and $\phi$ is azimuth angle) as: 
\begin{equation} 
  d(r, \theta, \phi) \propto \exp{\biggl (- \frac{{(\theta - \pi/2)}^2}{\sigma^2} \biggr ).}
\end{equation}
Here, $\sigma$ represents the characteristic angular width of the torus, describing the angular dispersion of the clumps around the equatorial plane.

The XCLUMPY model has been successfully applied to
spectral analysis of many AGN (e.g,
\citealt{2020ApJ...897....2T, 2021ApJ...906...84O,
  2021ApJ...913...17U, 2022ApJ...939...88I, 2023MNRAS.523.6239N}).
However, the original implementation of XCLUMPY, designed to apply to CCD
energy-resolution spectra, did not account for realistic
emission-line profiles (\citealt{1997PhRvA..56.4554H}), making it
impossible to reproduce the detailed Fe~K line profiles revealed by the
high-resolution spectra obtained with XRISM/Resolve.  In addition, it
has been recently suggested that the abundances of heavy elements such as Fe
and Ni in AGN tori may deviate from the solar composition adopted in the original XCLUMPY implementation
(Circinus galaxy; \citealt{2026arXiv260329748T}, and Centaurus~A; Nakatani et al., 2026, in preparation.).
To address these issues, \citet{2026arXiv260329748T} have developed an updated
version of the XCLUMPY model based on the MONACO framework
(\citealt{2016MNRAS.462.2366O}), applying to the Resolve spectrum of Circinus galaxy.

In this paper,we essentially adopt the same model, tuned to the spectral parameters of NGC~4388.
 The spectra were calculated over the 1$\sim$100 keV band using 20,000 logarithmically spaced energy grids, corresponding to an energy resolution of $\sim$ 1 eV around 6.4 keV. 
It includes realistic emission-line profiles and allows us to freely specify the elemental
abundances. 
For our model calculation, we assume the solar abundances
by \citet{2009LanB...4B..712L}.
For computational efficiency,  we calculate the
table model with the radiative transfer code
\textsc{SKIRT}(
\citealt{2015A&C.....9...20C}; \citealt{2020A&C....3100381C};
\citealt{2023A&A...674A.123V}; \citealt{2024A&A...688L..33V}), in
place of MONACO. We have confirmed that the differences
between SKIRT and MONACO do not affect our main conclusions
(\citealt{2026ApJ...997..352F}).
 In this model, we consider only the gas component and do not include dust grains.
To first order, the high-energy X-ray photons interact with individual atoms, regardless of them being locked up in solid dust grains. Typical dust features (e.g., X-ray absorption fine structures) are only visible with XRISM at very high S/N, e.g., in bright Galactic X-ray binaries.

The free parameters are the angular width of the torus $(\sigma)$ and the hydrogen
column density along the equatorial plane $({N^{\mathrm{Equ}}_{\mathrm{H}}})$.
Here, $\sigma$ is not the exact angular width of the torus,
but instead represents a characteristic angular width of the torus
(see Figure~\ref{figure:model_geomety} Right).
The abundances of Ca, Fe, and Ni are also treated as free parameters
$(Z(\mathrm{Ca}), Z(\mathrm{Fe}), Z(\mathrm{Ni}))$.

 NGC~4388 is known to host gas and dust along the polar direction of the torus, as revealed by observations with ALMA (\citealt{2021A&A...652A..98G, 2021A&A...652A..99A}) and VLTI (\citealt{2016ApJ...822..109A, 2019MNRAS.489.2177A}). 
Despite this evidence, here we adopt a spectral model that considers only reflection from the torus and the BLR in the X-ray band for simplicity.
Previous studies (e.g., \citealt{2019MNRAS.490.4344L, 2022MNRAS.512.2961M, 2026ApJ...997..352F}) have investigated the impact of polar dusty outflows on X-ray spectra. 
These works show that if Compton-thin gas is present along the polar direction, it can produce an excess in the spectrum below $\sim$3 keV due to scattering. 
In the present XRISM observation, however, the Resolve gate valve was closed, and therefore the spectral fitting was limited to energies above 3 keV. 
As a result, we could not well constrain the contribution from the polar gas. 
In this work, the scattering by such polar material (including highly ionized gas) is therefore approximated phenomenologically as a ``Thomson-scattered component'', which has the same spectral shape as that of the intrinsic component, following previous X-ray studies of obscured AGNs (e.g., \citealt{2016ApJS..225...14K, 2020ApJ...897....2T}).

\subsection{Broad Line Region}
\label{sec:BLRmodel}

Recent XRISM observations (e.g, \citealt{2024ApJ...973L..25X,
  2025PASJ...77S.209B, 2025ApJ...994L..10M, 2025ApJ...994L..13K})
indicate that the Fe K$\alpha$ emission line likely comprises multiple
components, including a narrow line from the torus and a broader
component that may arise in the BLR or inner disk.  Following the
approach of Nakatani et al. in preparation., we constructed a
reflection model to reproduce the broad emission-line
component originating from the BLR with the \textsc{SKIRT} code.  
The model assumes a geometrically thin,
disk-like structure with a uniform density 
(Figure \ref{figure:model_geomety} Left).
The free parameters are the half angular width of the disk and
 the hydrogen column density along the equatorial plane
$({N^{\mathrm{Equ}}_{\mathrm{H}}})$. 

In addition, this
model allows the abundances of Ca, Fe, and Ni to be varied so that they can be linked to the abundances of the XCLUMPY model (Section~\ref{sec:torus}).
 We do not model any kinematics in the SKIRT model, but rather smooth the SKIRT spectra in post-processing (gsmooth in XSPEC, see section~\ref{sec:fitting}).


\subsection{Photoionized Gas}
\label{sec:pionmodel}

For the ionized absorber and emitter, we calculate photoionization
models with \texttt{SPEX} (\citealt{1996uxsa.conf..411K}), generating
a grid of \texttt{pion} (\citealt{2016A&A...596A..65M,
  2015Natur.526..542M}) models covering a wide range of ionization
parameters and column densities.  The SED used as the input to
construct the \texttt{pion} model was generated by combining the
\texttt{comt}, \texttt{pow}, and multiple \texttt{etau} components.
This configuration reproduces a broadband AGN continuum, covering the
soft Comptonized emission at low energies and the power-law emission
at higher energies.  As a result, the illuminating spectrum is shaped
into a realistic form suitable for subsequent photoionization
calculations.

Since the intrinsic SED cannot be directly obtained from a type-2 AGN due to heavy obscuration, we adopted a representative Seyfert~1 SED based on NGC~5548 as a reference \citep{2015A&A...575A..22M}.
The free parameters are the ionization parameter $(\xi)$, outflow velocity $(v_{\mathrm{out}})$, hydrogen  column  density $(N_{\mathrm{H}})$, and velocity dispersion $(\sigma_{\mathrm{v}})$.

\section{Spectral Analysis and Results}
\label{sec:fitting}

In this section, we present the results obtained by modeling the
spectra from the XRISM/Resolve and NuSTAR/FPMs. The X-ray spectral
analysis is performed using XSPEC version: 12.14.1
(\citealt{1996ASPC..101...17A}). We adopt Cash statistic (C-stat;
\citealt{1979ApJ...228..939C}).

We evaluate the models using the Akaike Information Criterion (AIC; \citealt{1974ITAC...19..716A}), 
defined for Cash statistics as $\mathrm{AIC} = n \log{(C/n)} + 2p,$ where $n$ is the number of spectral bins, $C$ is the Cash statistic, and $p$ is the number of free parameters. 
Differences of $|\Delta \mathrm{AIC}| > 10$ are generally considered to indicate a strong preference for the model with the lower AIC, whereas values of $2 < |\Delta \mathrm{AIC}| < 10$ indicate a moderate (but not strong) preference for the model with the lower AIC.
Values of $|\Delta \mathrm{AIC}| < 2$ indicate that the additional model is unnecessary (e.g., \citealt{BurnhamAnderson2002}).

To model the broadband spectrum of the source, we fit the data in the
3--10 keV range for XRISM/Resolve, and in the 8--70 keV range for
NuSTAR/FPMA and FPMB with five models.
The statistics of each models are summarized in table~\ref{tab:model_comparison}.

\begin{deluxetable*}{lcccccc} \label{tab:model_comparison}
\tablewidth{\textwidth}
\tablecaption{Comparison of spectral models}
\tablehead{
Model & C-stat / dof & $\Delta C$ & $| \Delta \mathrm{AIC}|$  
}
\startdata
Model~1  (XCLUMPY only)& 2643.24 / 2518 & --- & --- \\
Model~2  (Model~1 + BLR) & 2559.40 / 2516 & $-83.84$ &  $79.84$ \\
Model~3  (Model~2 + XCLUMPY)& 2549.87 / 2514 & $-9.53$  &  $5.53$ \\
Model~4  (Model~3 + pion)& 2505.43 / 2510 & $-44.44$ &  $36.44$  \\
Model~5  (Model~4 + pion)& 2493.54 / 2507 & $-11.89$ &  $5.89$  \\
\enddata
\tablecomments{
Comparison of different spectral models. The values of $\Delta C$ and $|\Delta \mathrm{AIC}|$ are derived from a comparison between each model and the model shown in the line above.
The lowest model is the best-fit model adopted in this work, which provides the smallest residuals.}
\end{deluxetable*}

\subsection{Model~1: XCLUMPY}

We first tested a standard spectral model traditionally used for CCD-resolution data (e.g., \citealt{2020ApJ...897....2T, 2021ApJ...906...84O, 2021ApJ...913...17U, 2022ApJ...939...88I, 2023MNRAS.523.6239N}).
The X-ray spectral model is represented as follows in the XSPEC terminology:

\begin{eqnarray}\label{model1}
\mathrm{model_1}
&=& \textsf{const1*phabs} \nonumber\\
&*&  \textsf{(zphabs*cabs*zcutoffpl} \nonumber\\
&+& \textsf{const2*zcutoffpl + gsmooth*atable\{xclumpy.fits\}})\nonumber\\
\end{eqnarray}

\begin{enumerate}

\item The \textsf{const1} term is a cross-normalization constant to adjust small differences in the absolute flux calibration among different instruments. We set those of XRISM/Resolve to unity, and NuSTAR/FPMs to free. 
\item The \textsf{phabs} term represents the Galactic absorption, whose hydrogen column density is fixed at $2.87 \times {10}^{20}$ $\mathrm{cm}^{-2}$, a value estimated by the method of \citet{2013MNRAS.431..394W}. 
\item The first \textsf{zcutoffpl} component represents the direct X-ray emission.
The \textsf{zphabs} component accounts for the hydrogen column density
along the line of sight.
Note that the \textsf{cabs} model assumes free-electron scattering, whereas this work (SKIRT) assumes scattering by electrons bound to atoms. We confirm the difference in the total scattering cross section ($<10\%$) does not affect our fitting results \citep{2026ApJ...997..352F}. 
\item The second one is a scattered component from ionized gas in the polar region. The \textsf{const2} term denotes the scattered fraction. 
\item The last term represents the reflection spectra from the torus based on the XCLUMPY model
  (\citealt{2019ApJ...877...95T}).
 The photon index, cutoff energy, and normalization are linked to those in the \textsf{zcutoffpl} term.
  We first allowed the inclination angle to vary freely and found that the inclination angle and the angular width were strongly degenerate. Therefore, we fixed the inclination angle based on the values reported in \citet{2021ApJ...906...84O}.
 Because the cutoff energy is not well constrained by our data, 
we fix $E_{\mathrm{cut}} = 370 ~\mathrm{keV}$, the default value in the original XCLUMPY model \citep{2019ApJ...877...95T}, similar to those adopted in models of the X-ray background (see \citet{2014ApJ...786..104U} and references therein). We have confirmed that our results are hardly affected by adopting a lower value of $E_{\mathrm{cut}} = 200 ~\mathrm{keV}$ reported by \citet{2017ApJS..233...17R}.
We also link the Ca abundance to the Fe abundance in the fit, which cannot be well constrained from our data.
The \textsf{gsmooth} component represents the Doppler broadening assuming a Gaussian profile.

\end{enumerate}

First, we ignore the 6--8 keV
energy range of Resolve, to avoid the complexity introduced by the
presence of the Fe emission and absorption lines. This exclusion
is not expected to affect the continuum fit.
After determining the global spectral shape, we
fix $\log{N_{\mathrm{H}}^{\mathrm{Equ}}}$ to
$24.6^{+0.2}_{-0.4}~\mathrm{cm}^{-2}$, which is inferred in the
earlier step, to minimize coupling among spectral parameters, and perform a detailed
analysis including complex spectral features in the 6-8 keV band.  
This includes not only a detailed analysis of the Fe K$\alpha$ line
profile, but also an investigation of the absorption features using
the \texttt{pion} model.

After fixing $\log{N_{\mathrm{H}}^{\mathrm{Equ}}}$ to
$24.6~\mathrm{cm}^{-2}$, we fitted the overall spectrum, including the 6--8~keV band, obtaining a C-stat/d.o.f. of 2643.24/2518. 
We fined an Fe K$\alpha$ FWHM = $830^{+110}_{-110}$ km/s (Figure~\ref{figure:6-7keV} (a)).

\subsection{Model~2: XCLUMPY + BLR}
\label{sec:model2}
Significant residuals remain around the Fe K$\alpha$ line (Figure~\ref{figure:6-7keV} (a)); therefore, we next added a BLR reflection component to model~1.
The spectral model including the BLR component is expressed as follows in XSPEC terminology:

\begin{eqnarray}\label{model2}
\mathrm{model_2}
&=& \textsf{const1*phabs} \nonumber\\
&*&  \textsf{(zphabs*cabs*} \nonumber\\
&*& \textsf{(zcutoffpl + gsmooth*atable\{blr.fits\})}\nonumber\\
&+& \textsf{const2*zcutoffpl + gsmooth*atable\{xclumpy.fits\}})\nonumber\\
\end{eqnarray}

\begin{enumerate}
\item The \textsf{gsmooth*atable\{blr.fits\}} term represents the broadened emission lines from the BLR together with the associated reflected continuum (Section~\ref{sec:BLRmodel}). 
In our model, the BLR is treated as a uniform-density disk-like structure with a fixed angular width. Here we assume that the BLR does not intercept the
line of sight (i.e., $\sigma_{\mathrm{BLR}} < 20^{\circ}$ for $i = 70^{\circ}$);
otherwise, the total line-of-sight column density in the BLR and torus
required to explain the observed iron-K flux would substantially (by a factor of $>2$) exceed the observed absorption.
Since the BLR angular width cannot be well constrained from our data
beyond this constraint, we examined two representative values of the
BLR angular width (5$^\circ$ and 15$^\circ$).  We found that the fit
statistic was slightly better for the $15^\circ$ case, which is hence
adopted in our analysis.

The \textsf{zphabs*cabs} factor does only model extinction by the torus, as the BLR is assumed to have no line-of-sight extinction contribution for $i<7
5^{\circ}$. 
The photon index, cutoff energy, normalization and metal abundances are linked to those in the \textsf{zcutoffpl} term and XCLUMPY term.
\item The other terms are the same as Model~1.
\end{enumerate}

This model improves the fit to a C-stat/d.o.f. of 2559.40/2516, corresponding to $\Delta C = - 83.84$ and $|\Delta \mathrm{AIC}| = 79.84 > 10$.
This result indicates that the BLR component is strongly required to reproduce the X-ray spectrum of NGC~4388.
The narrow Fe~K$\alpha$ FWHM is $410^{+90}_{-70}$ km/s, and the broad FWHM is $3700^{+110}_{-850}$ km/s (Figure~\ref{figure:6-7keV}, (b)).

\subsection{Model~3: XCLUMPY (Two Components) + BLR}

However, some residuals remain in the red wing of the Fe K$\alpha$ line (Figure~\ref{figure:6-7keV}, (b)).
To investigate the possible origin of this feature, we examined an additional XCLUMPY component, which can be interpreted as emission from the inner edge region of the torus (Nakatani et al., in preparation), in the next model. 
The spectral model including the additional XCLUMPY component is expressed as follows in XSPEC terminology:

\begin{eqnarray}\label{model3}
\mathrm{model_3}
&=& \textsf{const1*phabs} \nonumber\\
&*&  \textsf{(zphabs*cabs} \nonumber\\
&*& \textsf{(zcutoffpl + gsmooth*atable\{blr.fits\})}\nonumber\\
&+& \textsf{const2*zcutoffpl}\nonumber\\
&+& \textsf{const3*gsmooth*atable\{xclumpy.fits\}}\nonumber\\
&+& \textsf{(1-const3)*gsmooth*atable\{xclumpy.fits\})}
\end{eqnarray}

\begin{enumerate}

\item The last two XCLUMPY terms represent the reflected spectrum from the dusty torus and its inner edge region, respectively. Since \textsf{const3} takes values between 0 and 1, the sum of the fluxes from the two components remains consistent with the case in which only a single XCLUMPY component is used.
All parameters in the XCLUMPY model are linked to each other. 
\item The other terms are the same as Model~2.

\end{enumerate}

We found that adding an additional XCLUMPY component improves the fit (Figure~\ref{figure:6-7keV}, (c)), yielding a decrease in the Cash statistic of 9.53  for two additional free parameters (C-stat/d.o.f. = 2549.87/2514), corresponding to $|\Delta \mathrm{AIC}| = 5.53 > 2$.
The derived narrow Fe K$\alpha$ component FWHMs are $260^{+110}_{-260}$ km/s and $1100^{+560}_{-390}$ km/s. The flux ratio of each narrow components are 0.57 : 0.43. The broad FWHM is $4800^{+2200}_{-1200}$ km/s.
Although $|\Delta \mathrm{AIC}| = 5.53$ does not indicate a strong preference for the additional XCLUMPY component, we include it as part of our best-fit model, motivated by the fact that the resulting three-component Fe K$\alpha$ emission scenario is consistent with that found in Centaurus~A (Nakatani et al., in preparation).
To confirm the need for this extra XCLUMPY component, we
perform an AIC test after including two \texttt{pion} components (see
Section~\ref{sec:model4} and \ref{sec:model5}) in the continuum
model. Comparing the fitting results between (1) Model~2 + pion + pion
and (2) Model 5, we obtain $|\Delta \mathrm{AIC}| = 8.49$,
indicating a high statistical significance for the presence of the
second velocity component in XCLUMPY.

We also tested a relativistic reflection model (RELXILL;
\citealt{2010MNRAS.409.1534D}) instead of the additional XCLUMPY
component.  However, the RELXILL model does not provide a
statistically significant improvement, with $|\Delta \mathrm{AIC}| <
2$.  This is likely because the large inclination angle causes the
relativistic disk line to be extremely broadened, making it
indistinguishable from the underlying continuum in the present data.

\subsection{Model~4: with Photoionized Absorption}
\label{sec:model4}

After fitting the broadband continuum and the Fe K$\alpha$ line with Model~3, residuals remain in the 6.4--7.0~keV band (Figure~\ref{figure:6-7keV} (c)), mainly associated with absorption lines.
To reproduce these features, we added a photoionized absorption model to Model~3.
The spectral model including the photoionized absorption model is expressed as follows in XSPEC terminology:

\begin{eqnarray}\label{model4}
\mathrm{model_4}
&=& \textsf{const1*phabs} \nonumber\\
&*&  \textsf{(zphabs*cabs*pion} \nonumber\\
&*& \textsf{(zcutoffpl + gsmooth*atable\{blr.fits\})}\nonumber\\
&+& \textsf{const2*zcutoffpl}\nonumber\\
&+& \textsf{const3*gsmooth*atable\{xclumpy.fits\}}\nonumber\\
&+& \textsf{(1-const3*gsmooth*atable\{xclumpy.fits\})}
\end{eqnarray}

\begin{enumerate}
\item The \textsf{pion} terms represent absorption by photoionized plasma. The model is calculated from the pion model (\citealt{2016A&A...596A..65M, 2015Natur.526..542M}) in SPEX (Section~\ref{sec:pionmodel}). 
\item The other terms are the same as Model~3.
\end{enumerate}

Including one \texttt{pion} zone significantly improves the fit in the  6.5--7.0 ~keV  (Figure~\ref{figure:6-7keV_de} (d)). 
Compared with the baseline model without photoionized 
absorption, the fit statistic decreases from 
$C = 2549.87$ for 2514 d.o.f. to 
$C = 2505.43$ for 2510 d.o.f., corresponding to an improvement of 
$\Delta C = - 44.44$ for 4 additional free parameters, i.e., $| \Delta \mathrm{AIC} | = 36.44 > 10 $.
We therefore conclude that a single highly ionized absorber is statistically required
to describe the Fe-K band. The best-fit parameters are tightly constrained to 
$\log N_{\mathrm{H}} = 22.1^{+0.2}_{-0.2}~\mathrm{cm}^{-2}$ and 
$\log \xi = 3.51^{+0.17}_{-0.13}$ erg cm s$^{-1}$, corresponding to a modest column density and a high ionization parameter that are consistent with 
Fe\,\textsc{xxv}/Fe\,\textsc{xxvi} absorption. 
The velocity dispersion and outflow velocity are  constrained to 
$\sigma_v = 160^{+100}_{-80}$ km s$^{-1}$ and $v_{\mathrm{out}} = 40^{+70}_{-70}$ km s$^{-1}$. 
Moreover, including this component does not affect the fit to the Fe K$\alpha$ line.

\subsection{Model~5: with Additional Photoionized Absorption (Best-Fit Model)}
\label{sec:model5}

Finally, to account for the residuals remaining around 6.4--6.5~keV (Fig~\ref{figure:6-7keV_de} (d)), where an additional absorption feature could in principle allow a stronger contribution from the BLR Fe K$\alpha$ emission while suppressing its high-energy wing,  we tested an additional \texttt{pion} component.
The spectral model including the second photoionized absorption model is expressed as follows in XSPEC terminology:

\begin{eqnarray}\label{model5}
\mathrm{model_5}
&=& \textsf{const1*phabs} \nonumber\\
&*&  \textsf{(const2*zphabs*cabs*pion*pion} \nonumber\\
&*& \textsf{(zcutoffpl + gsmooth*atable\{blr.fits\})}\nonumber\\
&+& \textsf{const3*zcutoffpl}\nonumber\\
&+& \textsf{const4*gsmooth*atable\{xclumpy.fits\}}\nonumber\\
&+& \textsf{(1-const4)*gsmooth*atable\{xclumpy.fits\})}
\end{eqnarray}

\begin{enumerate}
\item Two \texttt{pion}-absorption components are considered in this model. Since we could not constrain $\sigma_v$ of second absorption, we fixed it at 100 km s$^{-1}$ in the analysis. 
\item The other terms are the same as Model~4.
\end{enumerate}

Introducing the second absorption zone  improves the fit from
$C=2505.43$ for 2510 d.o.f. to $C=2493.54$ for 2507 d.o.f., corresponding to $\Delta C = - 11.89$ for three additional free parameters, i.e., $|\Delta\mathrm{AIC}| = 5.89 > 2$ (Figure~\ref{figure:6-7keV_de} (e)).
This indicates that a second \texttt{pion} component is required to reproduce the absorption features.
Similar ionized wind ($\log \xi \sim 2.5$ erg cm s$^{-1}$) components have been reported in well-studied sources such as NGC 3783 \citep{2025A&A...699A.228M} and NGC 3516 \citep{2025arXiv251207950J}, we included this component as part of the best-fit model in our analysis.
We obtained the second pion components parameters as $\log N_{\mathrm{H}} = 22.1^{+0.1}_{-0.1}~\mathrm{cm}^{-2}$,
$\log \xi = 2.61^{+0.10}_{-0.08}$ erg cm s$^{-1}$ and $v_{\mathrm{out}} = -100^{+80}_{-80}$ km s$^{-1}$ (inflow).

In addition, this component helps to reduce the residuals remaining in the red wing of the Fe K$\alpha$ line.
As a result, the line FWHM changes from those obtained with Models~3 and~4 to $290^{+70}_{-80}$ km s$^{-1}$, $1470^{+490}_{-340}$ km s$^{-1}$ (narrow components), and to $11100^{+3400}_{-3000}$ km s$^{-1}$ (broad component), respectively.
The flux ratio of each of the narrow components are 0.60 : 0.40.
The best-fit parameters are summarized in table \ref{table:broadbest}.

\begin{deluxetable*}{lcCcc}[htb]
  \tablecaption{Best fit parameters for the X-ray spectra of NGC 4388}
 \label{table:broadbest}
 \tablewidth{0pt}
 \tablehead{
 \colhead{Region} & \colhead{No.}&\colhead{Parameter} &\colhead{model~5 (final)} &\colhead{Units}
 }
 \startdata
 Torus  &(1) & $\log{N_\mathrm{H}^\mathrm{LOS}}$                             & $23.6^{+0.0}_{-0.1}$           & $\mathrm{cm}^{-2}$   \\
        &(2) & $\log{N_\mathrm{H}^\mathrm{Equ}}$                    & $24.6^a$&                  $\mathrm{cm}^{-2}$   \\
        &(3) & $\sigma$                                               & $12.5^{+2.5}_{-2.4}$                       & degree \\
        &(4) & $i$                                                              &   $70.0^a$                            & degree \\
        &(5) & $Z(\mathrm{Fe})$                                      & $1.94^{+0.21}_{-0.89}$                & solar \\
        &(6) & $Z(\mathrm{Ni})$                                      & $2.54^{+0.60}_{-0.94}$                & solar \\
        &(7) & $\mathrm{const3}$                               & $0.60^{+0.10}_{-0.12}$                &  \\
        &(8) & $\mathrm{FWHM_1(Fe)}$                  & $2.9^{+0.7}_{-0.8}\times10^2$         & km s$^{-1}$\\
        &(9) & $\mathrm{FWHM_2(Fe)}$                 & $1.5^{+0.5}_{-0.3}\times10^3$        & km s$^{-1}$\\
        &(10) & $\Gamma$                                             & $1.69^{+0.02}_{-0.03}$                &  \\
        &(11)& $E_{\mathrm{cut}}$                                               & $370^a$                               & keV  \\
        &(12)& $N_X$                                & $2.19^{+0.16}_{-0.16}\times{10^{-2}}$ &$\mathrm{photons}~\mathrm{keV}^{-1}~\mathrm{cm}^{-2}~\mathrm{s}^{-1}$\\
        &(13)& $f_{\mathrm{scat}}$                   & $1.77^{+0.20}_{-0.21}\times{10}^{-2}$ &   \\
 BLR    &(14)& $\log{N_\mathrm{H}^\mathrm{Equ}}$                           & $23.8^{+0.9}_{-0.2}$                  & $\mathrm{cm}^{-2}$   \\
        &(15)& $\mathrm{FWHM(Fe)}$ & $1.11^{+0.34}_{-0.30}\times10^4$             &km s$^{-1}$\\
 PION 1 &(16)& $v_{\mathrm{out}}$                                                  & $40^{+50}_{-40}$                   & km s$^{-1}$ \\ 
        &(17)& $\log{\xi}$                                                             & $3.50^{+0.09}_{-0.11}$                    & erg cm s$^{-1}$ \\
        &(18)& $\log{N_\mathrm{H}}$                                                    & $22.1^{+0.1}_{-0.1}$                       & $\mathrm{cm}^{-2}$   \\
        &(19)& $\sigma_v$                                                        & $160^{+60}_{-50}$                           & km s$^{-1}$ \\
PION 2  &(20)& $v_{\mathrm{out}}$                                                     & $-100^{+80}_{-80}$                                    & km s$^{-1}$ \\ 
        &(21)& $\log{\xi}$                                                             & $2.61^{+0.10}_{-0.08}$                                          & erg cm s$^{-1}$ \\
        &(22)& $\log{N_\mathrm{H}}$                                                   & $22.1^{+0.1}_{-0.1}$                                         & $\mathrm{cm}^{-2}$   \\
        &(23)& $\sigma_v$                                                            & 100$^a$                                                      & km s$^{-1}$ \\
others  &(24)& $C_{\mathrm{cross}}$                                & $0.95^{+0.01}_{-0.01}$                   & \\
        &    & $\mathrm{C-stat}/\mathrm{dof}$                             & $2493.54/2507$                              &  \\
 \enddata
 \tablecomments{
 (1) Hydrogen column density along the line of sight. (2) Torus hydrogen column density along the equatorial plane.
 (3) Torus angular width. (4) Inclination angle. (5) Abundance of Fe relative to hydrogen. (6) Abundance of Ni relative to hydrogen. (7) Flux ratio between two narrow components. (8) FWHM of the narrow Fe K$\alpha$ component. (9) FWHM of the intermediate Fe K$\alpha$ component. (10) Photon index. (11) Cutoff energy. (12) Normalization of the intrinsic power law component at 1 keV. (13) Scattering fraction. (14) BLR hydrogen column density along the equatorial plane. (15) FWHM of the broad Fe K$\alpha$ component. (16) Outflow velocity.(17) Ionization parameter. (18) Hydrogen column density of the outflow. (19) Velocity dispersion. (20)-(23) Second pion component. (24) Cross-normalization constant.
 \tablenotetext{a}{The parameter is fixed.}}
\end{deluxetable*}


\begin{figure*}[htb]
\centering
\includegraphics[width=0.8\textwidth]{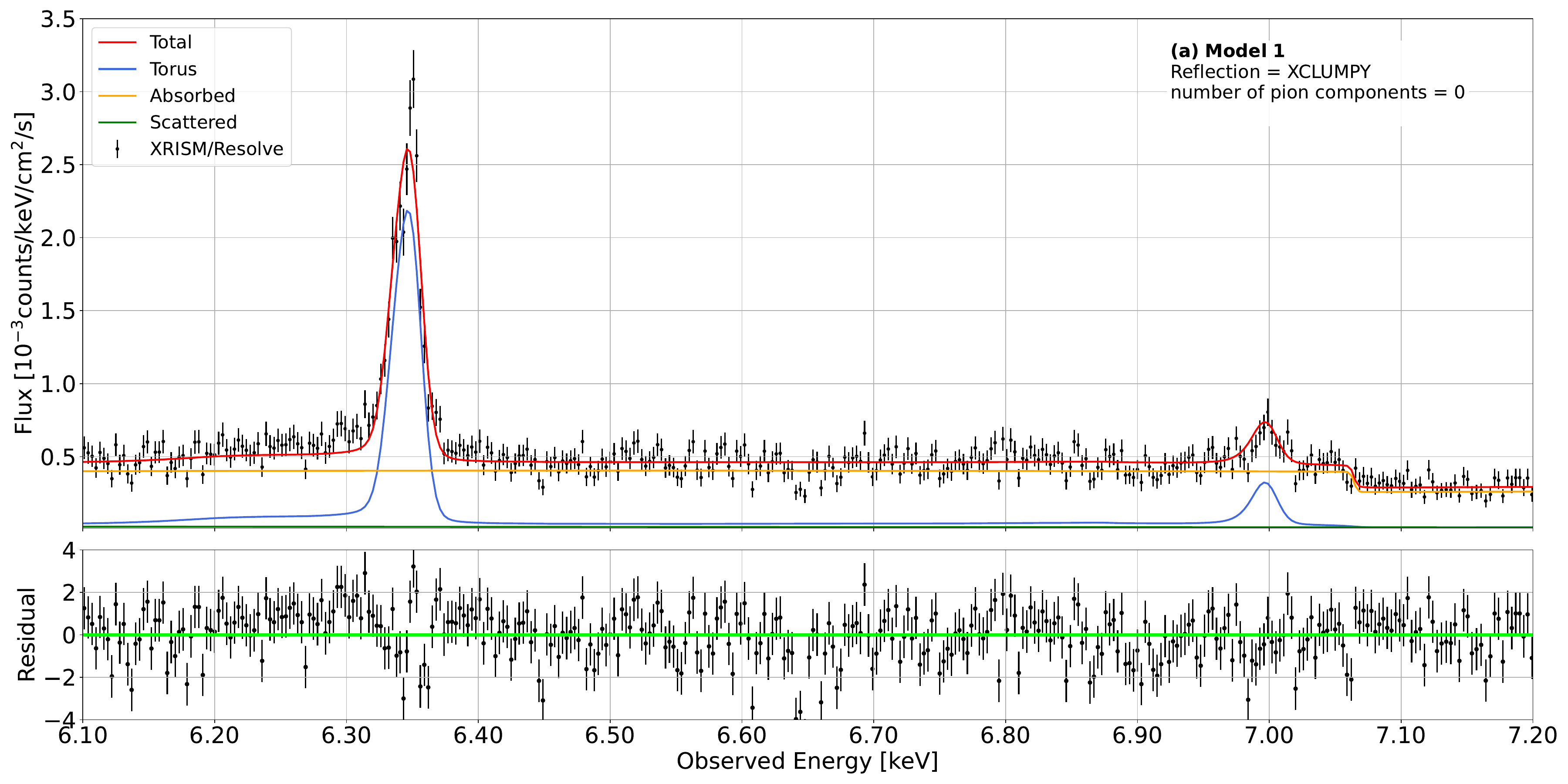}\\
\includegraphics[width=0.8\textwidth]{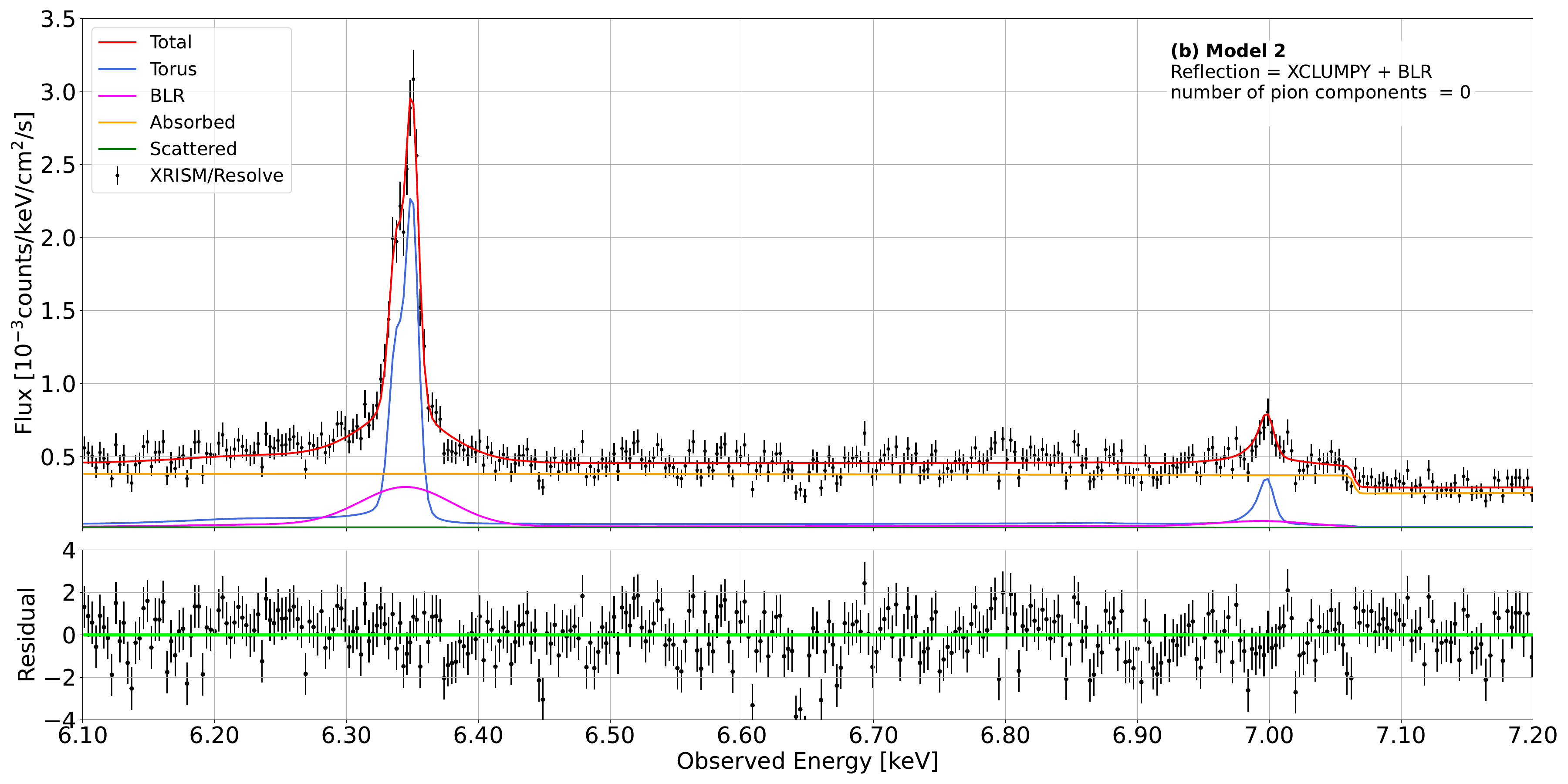}\\
\includegraphics[width=0.8\textwidth]{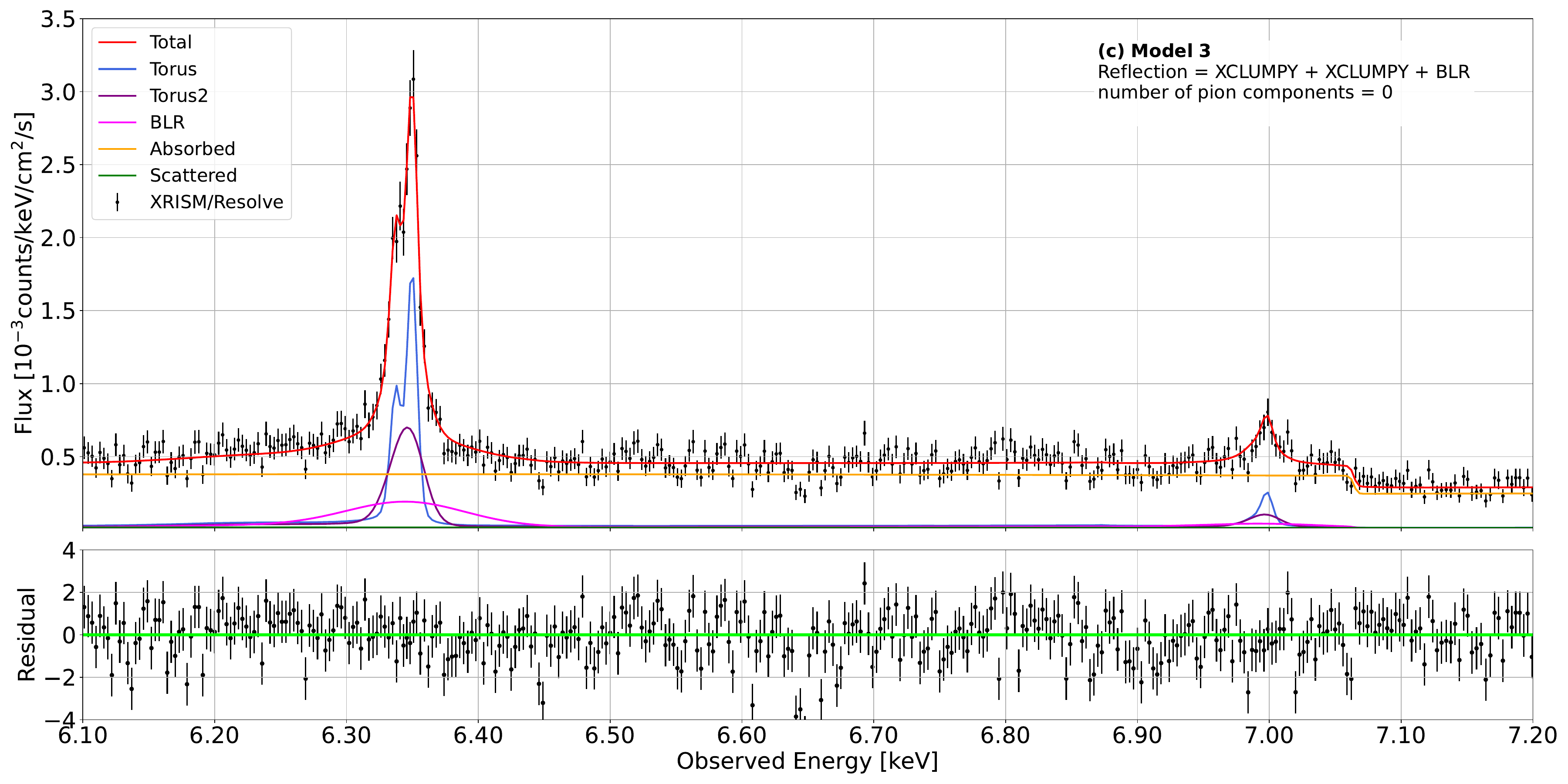}
\caption{
XRISM/Resolve spectrum of the Fe K region (black points) with the model (red) and its components.
(a) model~1; XCLUMPY only,
(b) model~2; XCLUMPY + BLR,
(c) model~3; XCLUMPY + XCLUMPY + BLR.
}
\label{figure:6-7keV}
\end{figure*}

\begin{figure*}[htb]
\centering
\includegraphics[width=0.8\textwidth]{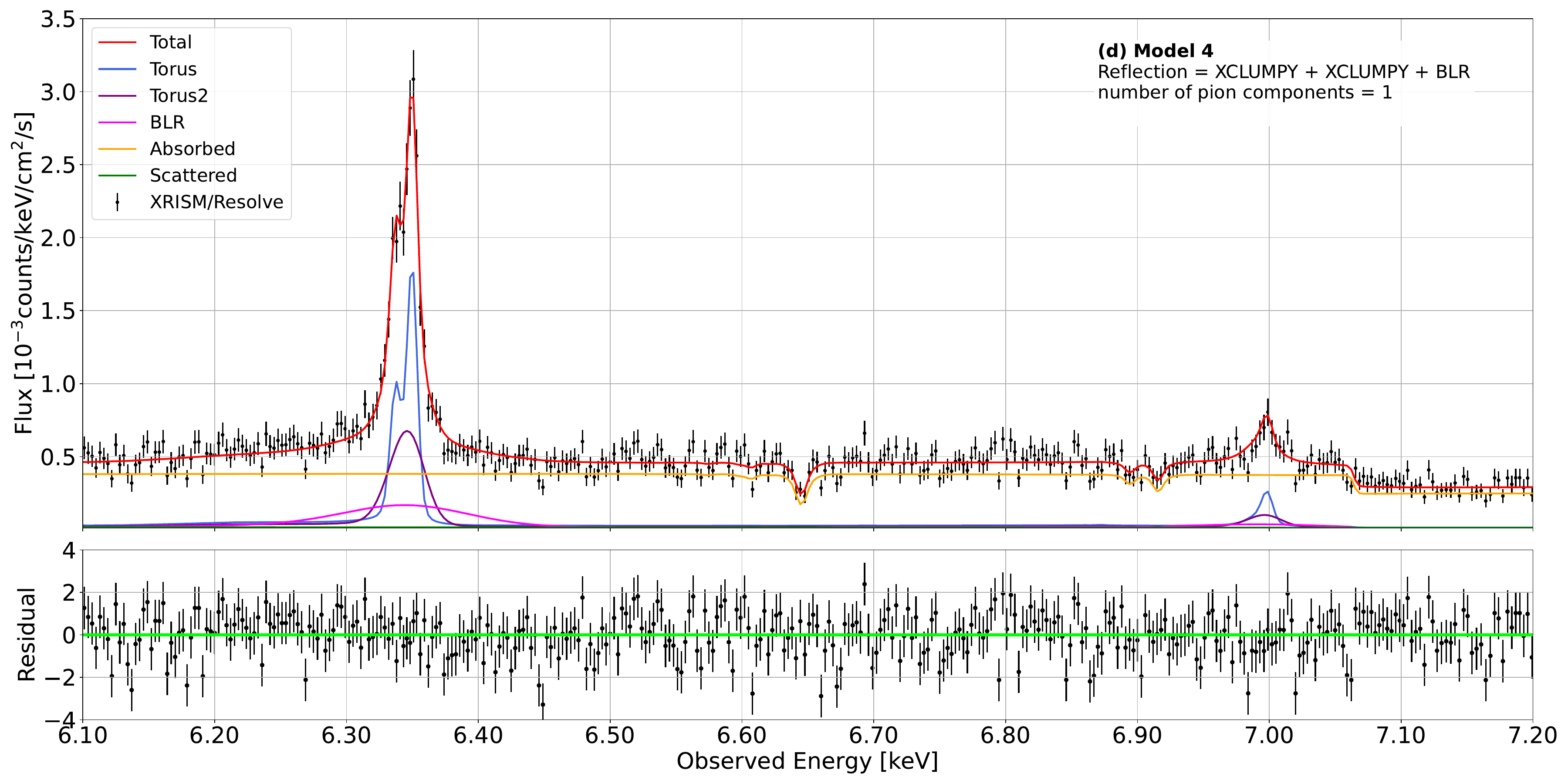}\\
\includegraphics[width=0.8\textwidth]{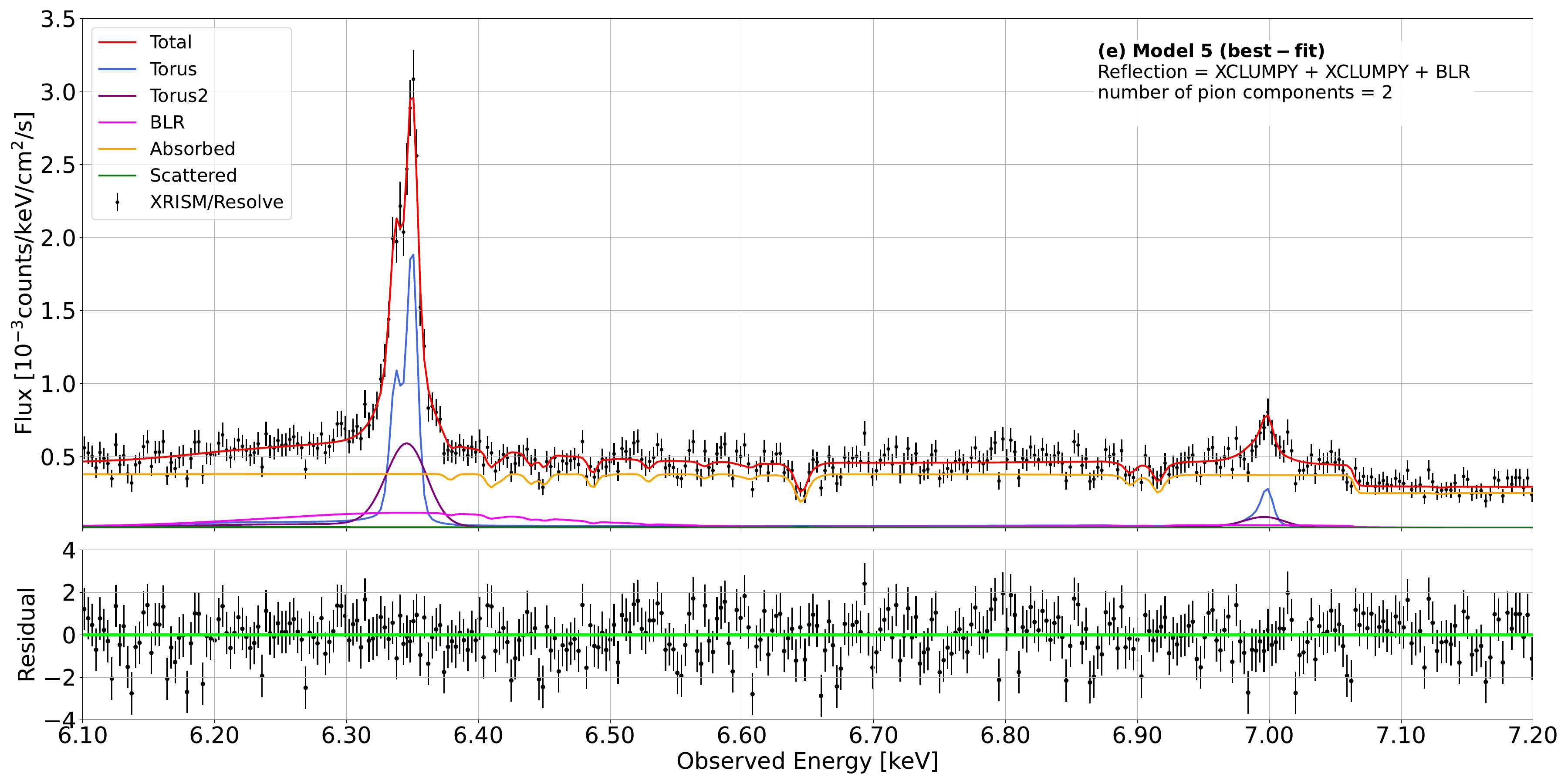}
\caption{
(continued)
(d) model~4; XCLUMPY + XCLUMPY + BLR + pion,
(e) model~5; XCLUMPY + XCLUMPY + BLR + pion  + pion (final).
}
\label{figure:6-7keV_de}
\end{figure*}

\begin{figure*}[htb]
\centering
\includegraphics[width=0.9\textwidth]{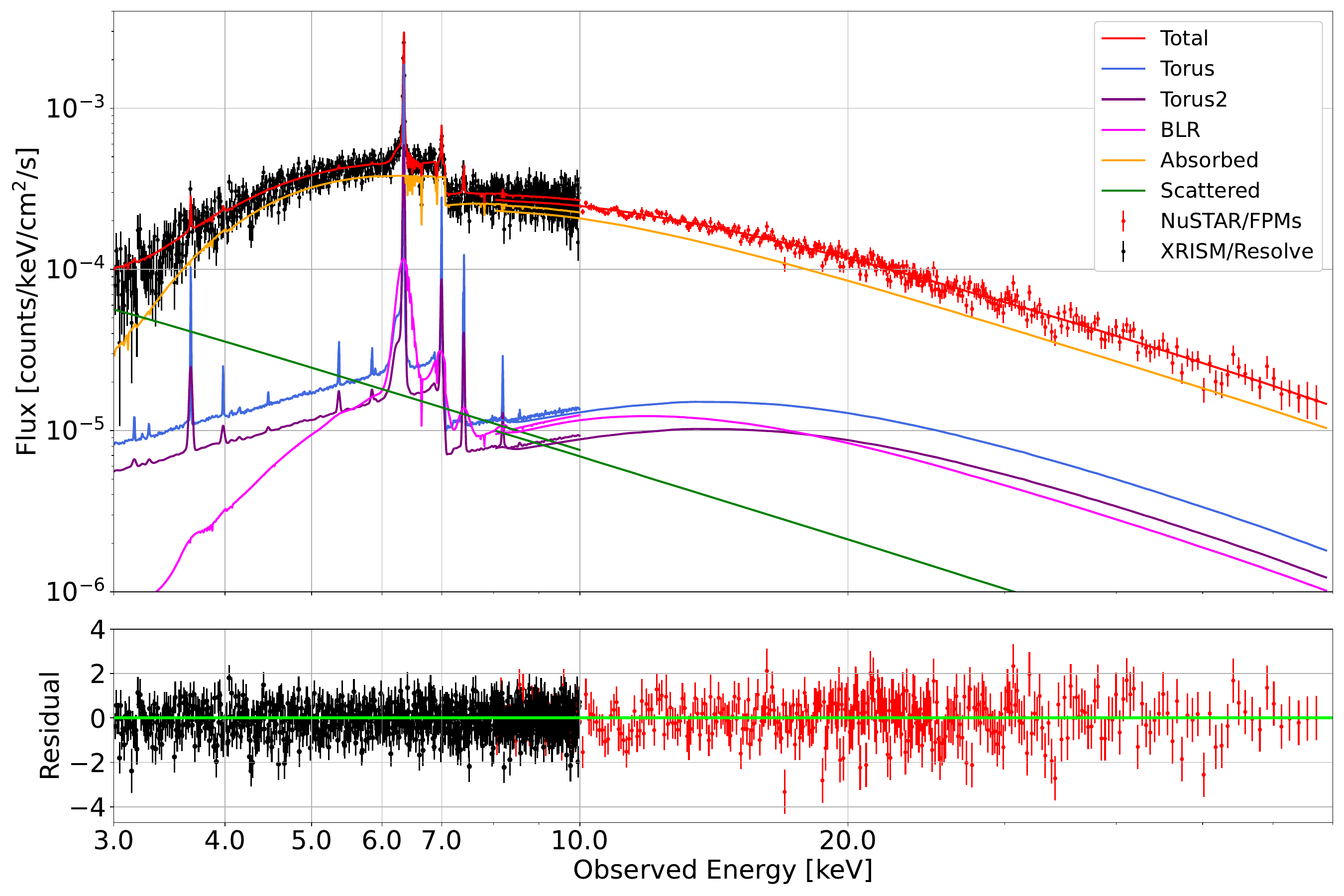}
\includegraphics[width=0.9\textwidth]{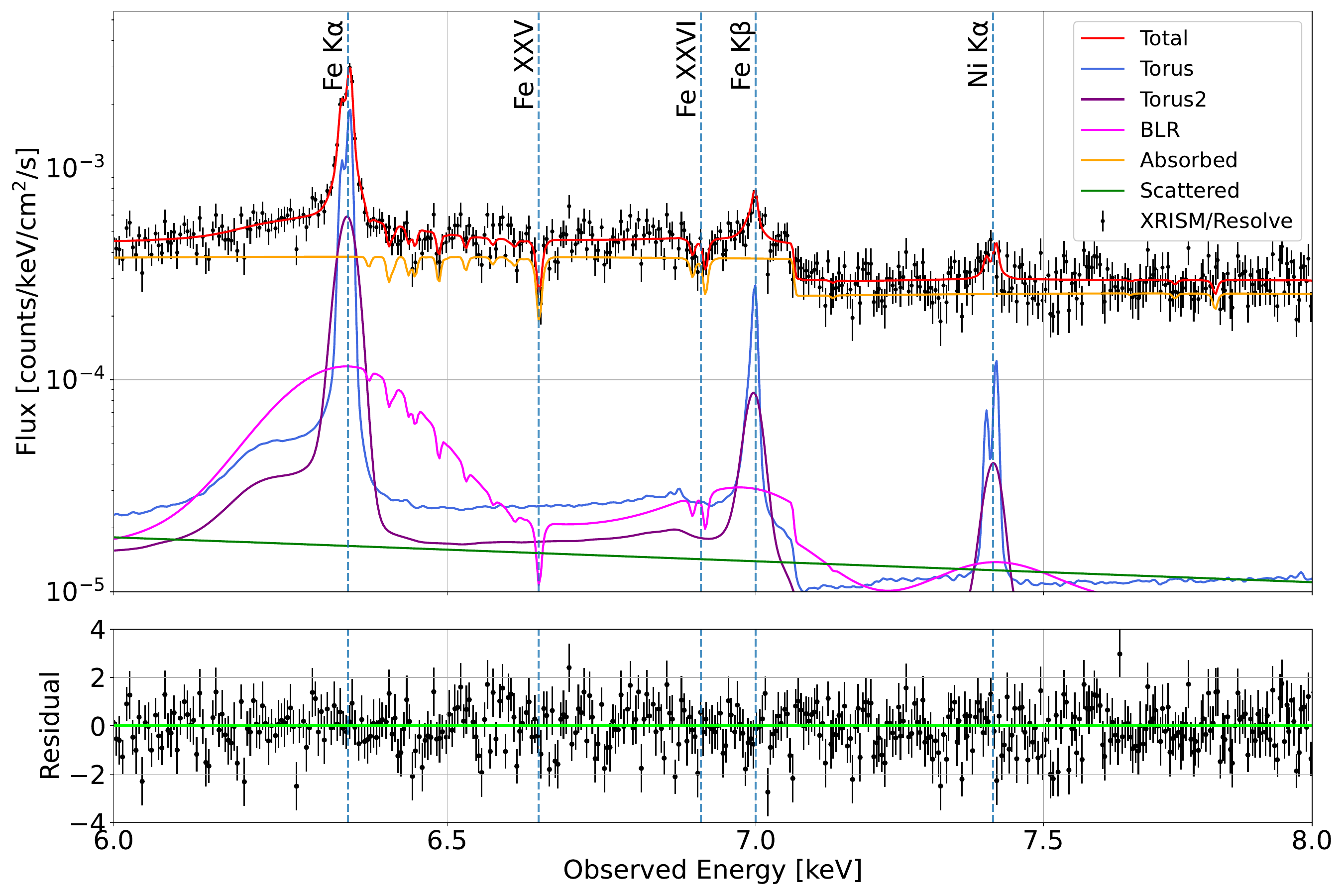}
\caption{Broad-band (top) and Fe K-band (bottom) spectral fits to the XRISM/Resolve (black) and NuSTAR/FPMA+FPMB (red) data.
The total best-fit model is shown in red, and the individual spectral components are plotted separately:  the torus-reflection components (blue and purple), the BLR-reflection component (magenta), the absorbed power-law continuum (orange), and the scattered power-law component (green). 
Prominent fluorescent and absorption lines in the Fe K complex (Fe K$\alpha$, Fe XXIV–XXV, Fe XXVI Ly$\alpha$, Fe K$\beta$) as well as the Ni K$\alpha$ line are labeled in the lower panel.
Residuals (data minus model, normalized by the statistical uncertainties) are shown beneath each spectrum.
The broad-band panel demonstrates that the combined torus and BLR reflection components reproduce the continuum curvature and the strong Fe K$\alpha$ emission, while the zoomed-in Fe K-band panel highlights the detailed line structure resolved by XRISM/Resolve.}
\label{fig:broadband_combined}
\end{figure*}

\begin{figure*}[htb]
\centering
\includegraphics[width=1\textwidth]{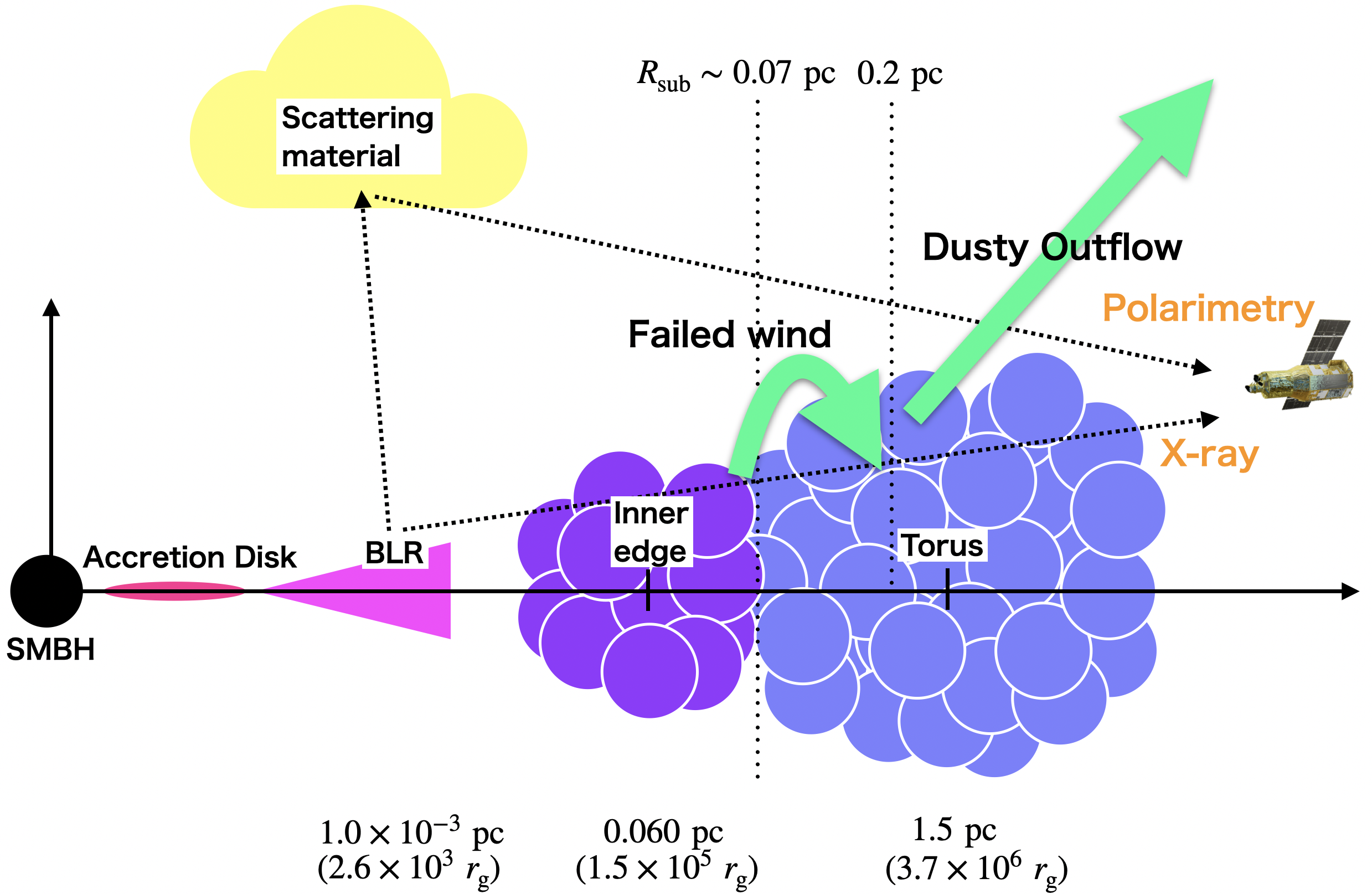}
\caption{
Schematic illustration of the circumnuclear environment of NGC 4388 inferred from the joint XRISM/Resolve and NuSTAR/FPMs observations.
The central black hole is surrounded by the broad-line region and an obscuring torus, whose geometry and composition are constrained by the spectral decomposition.
An ionized absorber detected in the Resolve data exhibits low outflow velocity indicating that the gas is gravitationally bound rather than escaping.
This component is naturally interpreted as a failed wind, launched from the irradiated inner surface of the torus and decelerated as dust sublimates at larger radii.
The viewing angle places our line of sight close to the torus rim, allowing simultaneous access to the reprocessed emission and the failed-wind absorption.
\label{Figure:NGC4388}}
\end{figure*}

\section{Discussion}
\label{sec:discussion}

We have analyzed the high–resolution X-ray spectrum of NGC~4388
obtained with XRISM/Resolve in the 3--10 keV band, 
jointly with simultaneous NuSTAR spectra that cover the 8--70 keV band.
Since NGC 4388 shows the brightest Fe K line among Compton-thin absorbed AGNs (except for the radio galaxy Centaurus A), these data allow us to investigate the structure of an obscured AGN in unprecedented detail.

To fit the broadband spectra, we have employed physically motivated
models for the reprocessed emission which includes the fluorescent lines:
an updated version of XCLUMPY (\citealt{2019ApJ...877...95T}) for the torus and a
newly developed model by Nakatani et al. (2026, in preparation) for the BLR. 
These models are calculated
with the SKIRT code (\citealt{2023A&A...674A.123V}; \citealt{2024A&A...688L..33V}), where realistic Fe K line profiles based
on ground measurements (\citealt{1997PhRvA..56.4554H}) as well as variable metal abundances
are taken into account. We note that the inclusion of the reflection
continuum for the latter component, which has been ignored in previous
works, is important to correctly model the broadband spectrum.
Now we have a more complete view on the circumnuclear environment by including a model for the torus and the BLR. 
The absorption features by ionized
gas have been modeled by the pion code, following previous studies
based on XRISM data (\citealt{2024ApJ...973L..25X, 2025ApJ...988L..54X, 2025ApJ...994L..10M, 2025A&A...699A.228M, 2025arXiv251207950J} ).

Our best-fit parameters are summarized in table \ref{table:broadbest},
and the  inferred geometry  is shown in Figure \ref{Figure:NGC4388}. We confirm that the basic
parameters of the continuum emission (photon index and line-of-sight
absorption column density) and those of the torus parameters in
XCLUMPY (the torus angular width, inclination, and the equatorial
column density) are mostly consistent with earlier works based on the
CCD and NuSTAR data (e.g.,\citealt{2021ApJ...906...84O}).
Our spectral analysis yields the Ni/Fe abundance ratio of 
$\sim\ 1.3$ solar.
It is noteworthy that the super-solar ratio of Ni/Fe is also found in
      Circinus galaxy (\citealt{2026arXiv260329748T}) and Centaurus A (Nakatani et al., in preparation.).
We leave it to a future work to constrain the metal enrichment history
of the nucleus in NGC 4388 by combining the abundances of lighter
elements as done in \citet{2026arXiv260329748T}. In the following, we discuss the
structure of the cirumnuclear material of NGC 4388, mainly focusing on
the widths of Fe K line components (Section~\ref{sec:torusBLR}) and the parameters of
ionized gas (Section~\ref{sec:ionizedgas}).

\subsection{Torus and BLR}
\label{sec:torusBLR}

Our analysis reveals that the Fe~K$\alpha$ emission line profile in
NGC 4388 is best described by a composition of three components
convolved with Gaussians of different line widths
(Figure \ref{figure:6-7keV_de} (e)): narrow (with an FWHM of
$290^{+70}_{-80}$ km s$^{-1}$ in Doppler velocity), intermediate
($1470^{+490}_{-340}$ km s$^{-1}$), and broad ($11100^{+3400}_{-3000}$
km s$^{-1}$) . Under the assumption of Keplerian rotation,
these velocity widths correspond to radii of 3.7$^{+3.2}_{-1.3} \times 10^{6}$
r$_{\rm g}$ (r$_g \equiv \mathrm{GM/c^2}$ is a gravitational radius where G is
the gravitational constant, M is the black hole mass, and c is the
speed of light), $1.5^{+1.0}_{-0.7} \times 10^{5} $ r$_{g}$, and
$2.6^{+2.3}_{-1.1} \times 10^{3}$ r$_g$, respectively, for an inclination of 70
degree (Table~\ref{table:broadbest}).  Adopting the black hole mass of $8.5\times10^6 ~M_{\odot}$,
these values are converted to $r \sim 1.5$~pc, $r \sim 0.060$~pc, and
$1.0\times10^{-3}$ pc, respectively. 
\citet{2024ApJ...966...57G} find evidence for a reverberation lag of 
$t = 16$ days, corresponding to a radial scale of 
$r = 0.014$ pc ($\sim3.4 \times 10^{4}\,\mathrm{r_g}$). 
This value is comparable to the location of the intermediate FWHM component derived in our analysis 
($\sim0.060$ pc), suggesting that both studies consistently trace gas at a similar characteristic distance from the central engine. 
A relativistically broadened
Fe K line from the accretion disk close to the SMBH ($<$ 100~ r$_{\rm g}$) is
not significantly detected, confirming the result by \citet{2023MNRAS.522..394Y}.  
The reason is unclear; the high inclination of this system might make it difficult
to be observed if e.g., the disk were warped, partially blocking its
innermost region.

To identify the origin of the Fe K line emitting regions, we compare these
locations with that of the ``dusty'' torus. 
\citet{2008ApJ...685..147N, 2008ApJ...685..160N} showed that the
dust sublimation radius $R_{\mathrm{sub}}$ can be obtained by the
following equation.
\begin{equation}
R_{\rm sub} = 0.4 \left( \frac{L_{\rm bol}}{10^{45}\ {\rm erg\ s^{-1}}} \right)^{1/2} \left( \frac{1500\ {\rm K}}{T_{\rm sub}} \right)^{2.6}{\rm \,pc}
\end{equation}
Here, $L_\mathrm{bol}$ denotes the bolometric luminosity of the AGN and $T_{\rm
  sub}$ is the dust sublimation temperature.  Assuming $T_{\rm sub} =
1500$~K and adopting the bolometric luminosity obtained from the
broadband spectral analysis ($L_{2-10~\mathrm{keV}}\sim 1.4\times10^{43}$ erg 
s$^{-1}$ and $L_{\mathrm{bol}}=20\times L_{\mathrm{2-10~\mathrm{keV}}}$), we
estimate the dust sublimation radius of NGC 4388 to be $R_{\rm
  sub} \sim 0.21$~pc. \citet{2007A&A...476..713K} showed that the
inner radius of the torus inferred from near-infrared reverberation
mapping is typically a factor of $\sim3$ smaller than the sublimation
radius estimated in this manner.  This discrepancy can be explained by
the anisotropic radiation field of the accretion disk
(\citealt{2010ApJ...724L.183K}).  Applying this correction factor, we
infer that the inner radius of the dusty torus in NGC 4388 is
approximately $0.07$~pc.

This suggests that the narrow component (FWHM $\sim
290\ \mathrm{km\ s^{-1}}$; $r \sim 1.5 $~pc) originates in the dusty
torus, located at radii beyond the dust sublimation radius.
Furthermore, the inferred location of the intermediate component (FWHM $\sim
1470\ \mathrm{km\ s^{-1}}$; $r \sim 0.060$~pc) lies inside the dust
sublimation radius.  
This implies that this component may be partially
associated with dust-free gas located closer to the SMBH than the
dusty torus. This interpretation is in line with the arguments
suggesting that a significant fraction of the Fe K$\alpha$ line arises
from regions closer to the SMBH than the dusty torus (e.g., 
\citealt{2015ApJ...802...98M, 2015ApJ...812..113G, 2021ApJ...913...17U, 2024MNRAS.532..666M}).

The line width of the broad Fe K$\alpha$ component (FWHM $\sim
10^4\ \mathrm{km\ s^{-1}}$) is within a typical range of optical
``broad lines'' in AGNs (e.g., \citealt{2022ApJS..261....5M}), supporting our interpretation that it
is likely originates from the same region as in the optical
BLR. Following the case of Centaurus A (\citealt{2025PASJ...77S.209B}; Nakatani et
al., in preparation.), this demonstrates the power of X-ray observations that
can directly probe  the structure inside the dust torus even in
type-2 AGNs.

In NGC 4388, the broad component of H$\alpha$ is measured
through spectropolarimetry \citep{2016MNRAS.461.1387R},
and shows an FWHM $\sim 4500 \pm 1400~\mathrm{km\,s^{-1}}$.
Notably, this is smaller than the line width of the broad component of
Fe K$\alpha$ ($11100^{+3400}_{-3000}$
km s$^{-1}$), although still (barely) consistent within
the errors. This situation might be different from the Seyfert 1.5
galaxy NGC 4151, where the line widths of H$\alpha$ from the optical
BLR and Fe K$\alpha$ line from the X-ray BLR
\citep{2024ApJ...973L..25X} simultaneously observed are similar to each
other (Noda et al. in preparation.). The difference between the Fe
K$\alpha$ and H$\alpha $ line widths in NGC 4388 could be understood as an inclination effect; in X-rays the BLR is directly edge-on, whereas in the optical band, the direct BLR is fully obscured. Yet, the optical BLR is seen face-on by the gas in the polar direction, and this face-on view is scattered to the observer in polarized optical light.  This explains our results if the velocity field of the BLR is dominated by motions parallel to the equatorial plane (i.e., Keplerian rotation) rather than that perpendicular to it.
Thus, high resolution X-ray spectroscopy of obscured AGNs has a
potential to perform ``tomography'' of the BLR, which is useful
to constrain its theoretical models (e.g. \citealt{2011A&A...525L...8C}). 

It is remarkable that the type 2 radio galaxy Centaurus A also shows a
similar Fe K$\alpha$ profile that is well represented by the three
line-width components (Nakatani et al., in preparation.). These facts suggest
that the nuclear structures we have revealed may be common among AGNs
in a wide range of Eddington ratio. Detailed comparison of Fe
K$\alpha$ profile in the Resolve spectra among all the PV AGN targets
will be presented in a forthcoming paper (Ueda et al., in preparation.)

\subsection{Low Velocity Ionized Outflow}
\label{sec:ionizedgas}

We have detected Fe\,\textsc{xxvi} Ly$\alpha$ and Fe\,\textsc{xxv} absorption lines, indicating the presence of a highly ionized absorber with 
$\log \xi = 3.50\ \mathrm{erg\,cm\,s^{-1}}$, 
$\log N_{\mathrm{H}} = 22.1\ \mathrm{cm^{-2}}$, 
an outflow velocity of $v_{\mathrm{out}} = 40\ \mathrm{km\,s^{-1}}$, 
and a velocity dispersion of $\sigma_{v} = 160\ \mathrm{km\,s^{-1}}$. 
These values are consistent with the NICER observation reported by \citet{2019ApJ...884..106M}.
The ionization parameter of the absorber is defined as
\begin{equation}
\xi = \frac{L_{\mathrm{ion}}}{n r^{2}},
\end{equation}
where $L_{\mathrm{ion}}$ is the ionizing luminosity, $n$ is the gas density, 
and $r$ is the distance from the central SMBH.
Using the relation between the gas density and its column density,
$N_{\mathrm{H}} = n\,\Delta r$, and , assuming the absorber does not extend over a wide range of radii ($\Delta r < r$, e.g.\ \citealt{2012ApJ...753...75C, 2013MNRAS.430.1102T}), such that a single-zone photo-ionisation description remains meaningful, we obtain an upper limit on the distance
    \begin{equation}
    r < \frac{L_{\mathrm{ion}}}{\xi\, N_{\mathrm{H}}}.
    \end{equation}
This inequality provides an upper limit on the distance of the ionized absorber, 
$r < 0.23 $~pc ($5.8\times10^{5}~r_g$), by assuming ${L_{\mathrm{ion}} \sim 0.1\,L_{\mathrm{bol}}}$  
(\citealt{2007MNRAS.379.1359M, 2009MNRAS.392.1124V, 2010A&A...512A..34L}). 
This radius is almost consistent with the value inferred from the Chandra/HETG observations 
(\citealt{2024ApJ...966...57G}).
At this radius, the escape velocity from the central black hole 
($M_{\mathrm{BH}} \sim 8.5\times10^{6}\ M_\odot$; \citealt{2011ApJ...727...20K}) is 
$v_{\mathrm{esc}} \sim 560~ \mathrm{km\,s^{-1}}$, 
which is more than an order of magnitude larger than the observed outflow velocity 
($v_{\mathrm{out}} \sim 40$~km\,s$^{-1}$).

Given the high ionization state of the absorber, we first examine
the possibility that it originates at the base of a  Ultra Fast Outflow (UFO)  or a failed
UFO wind (e.g., \citealt{2015MNRAS.446..663H, 2021MNRAS.503.1442M}).
However, the relatively small velocity dispersion
($\sigma_v \sim 160~\mathrm{km\,s^{-1}}$), 
which is likely to reflect the local gravitational potential, 
is difficult to reconcile with the small launching radius of a UFO inferred from its high bulk velocity.
Thus, we suggest that the absorber is associated with a failed outflow
arising at much large radii close to the dusty torus region (with a
dust sublimation radius of $\sim$ 0.07 pc).

It is theoretically predicted that at such radii radiation pressure on
dust can lift gas off the torus surface, producing a dusty outflow
(e.g., \citealt{2012ApJ...758...66W, 2016ApJ...828L..19W, 2023ApJ...950...72K}).  As the gas rises and penetrates inside
the sublimation region, dust grains directly irradiated by the central
emission are sublimated and the opacity drops, causing a rapid
decrease of the radiative force. Since the resulting dust-free, highly
ionized gas remains gravitationally bound at $r \lesssim 0.2$~pc
it eventually fall back toward the disk, forming a fountain-like (e.g. \citealt{2012ApJ...758...66W, 2016ApJ...828L..19W, 2022ApJ...925...55O}), failed wind (Figure~\ref{Figure:NGC4388}).
Such dynamical structures have been observationally supported in the Circinus galaxy with ALMA by \citet{2018ApJ...867...48I,2023Sci...382..554I}, who
show that the line profile of [C~I] $^3{\rm P}_1 - ^3{\rm P}_0$ line is consistent with a fountain flow.
A part of outflow where dusts survive can be escaped from the system, producing a large-sale dusty polar outflow.
In fact, polar extended emission is also observed in NGC 4388  (\citealt{2016ApJ...822..109A, 2019MNRAS.489.2177A}).
Taken together with the presence of a slow, highly ionized failed wind
at sub-parsec scales, these results suggest that a dynamic
mechanism—rather than a static configuration—plays a key role in
shaping the circumnuclear structure of the AGN, including the dusty
torus.

For the additional pion components, the derived distance from the SMBH is less than 2 pc. The upper limit on the outflow velocity ($\sim-20\ \mathrm{km\ s^{-1}}$) is also lower than the escape velocity at 2 pc ($\sim190\ \mathrm{km\ s^{-1}}$).
This second ionized component may share a similar origin to the first component discussed later, or alternatively may be associated with material inflowing toward the SMBH.

\section{Conclusion}

We have performed a broadband X-ray spectral analysis of the Compton-thin Seyfert 2 galaxy NGC 4388 using XRISM/Resolve, which provides the
highest energy resolution to date, together with simultaneous NuSTAR/FPMs
data. Figure~\ref{Figure:NGC4388} illustrates the schematic view of
the circumnuclear structures of NGC 4388 derived from our work.

\vspace{0.5\baselineskip}

\begin{enumerate}

\item To model the broadband spectra in a physically motivated manner,
  we employ an updated version of the XCLUMPY mode and a BLR model
  with disk-like geometry, both calculated with the \textsc{SKIRT}
  code. The intrinsic line profile of Fe K lines and variable metal
  abundances are taken into account. Furthermore, we model the
  absorption features by ionized gas using the \textsc{pion} code.

\item The torus parameters (an angular width of 12.6$^{\circ}$, an inclination of 70$^{\circ}$ , an equatorial hydrogen column density of 10$^{24.6}~\mathrm{cm}^{-2}$) are found to be consistent with previous results analyzing the broadband spectra with the XCLUMPY model.

\item The Ni/Fe abundance ratio is found to be super-solar (1.3$^{+0.5}_{-0.3}$ solar), similar to the case of Circinus galaxy and Centaurus A. 
  
\item The Fe K$\alpha$ fluorescence line is well described with three different velocity components with a FWHM of $\sim$ 290 km s$^{-1}$, $\sim$ 1470 km s$^{-1}$, and $\sim$11100 km s$^{-1}$. We
  interpret that they originate from the dusty torus, its inner edge
  region, and the BLR, respectively.  A comparison with the dust sublimation radius suggests that some fraction of the inner edge region lies interior to it and is composed of dust-free gas..  The line width of the BLR component is
  larger than that of H$\alpha$ in optical polarized light,
  implying that the velocity distribution of the BLR is dominated by
  motion parallel to the equatorial plane (i.e., Keplerian motion).

\item The absorption line features in the 6--8 keV band are mainly reproduced
  with a highly-ionized, slowly-moving photoionized gas characterized
  by $\log{\xi} \sim 3.50~\mathrm{erg\ cm\ s^{-1}}$,
  $\log{N_{\mathrm{H}}} \sim 22.1~\mathrm{cm^{-2}}$, $v_{\mathrm{out}}
  \sim 40\ \mathrm{km\ s^{-1}}$, and $\sigma_v \sim
  160\ \mathrm{km\ s^{-1}}$. We infer that it is a failed wind
  launched around the dust sublimation radius ($\sim$0.07 pc),
  consistent with a radiation-driven fountain flow as theoretically
  predicted.

\item We also detect another ionized component characterized by $\log N_{\mathrm{H}} = 22.1^{+0.1}_{-0.1}~\mathrm{cm}^{-2}$, $\log \xi = 2.61^{+0.10}_{-0.08}$ erg cm s$^{-1}$ and $v_{\mathrm{out}} = -100^{+80}_{-80}$ km s$^{-1}$ (inflow).
This component may share a similar origin to the first component, or alternatively be associated with material inflowing toward the SMBH.
\end{enumerate}

This work was supported by the Japan Society for the Promotion of Science (JSPS) KAKENHI grant number 20H01946 (Y.U.), 24K17104 (S.O.) and 21K13958 (M.M.).
Y.U. acknowledges the support from the Kyoto University Foundation.
LG acknowledges support from the Canadian Space Agency grant 25EXPRSM1.
This work made use of the JAXA Supercomputer System Generation 3 (JSS3).
This work was also supported by Yamada Science Foundation. 
 This research has made use of data and/or software provided by the High Energy Astrophysics Science Archive Research Center (HEASARC), which is a service of the Astrophysics Science Division at NASA/GSFC and the High Energy Astrophysics Division of the Smithsonian Astrophysical Observatory. This research has also made use of the NASA/IPAC Extragalactic Database (NED), which is operated by the Jet Propulsion Laboratory, California Institute of Technology, under contract with the National Aeronautics and Space Administration.
Facilities:  XRISM (201063010),
NuSTAR (60061228002, 60061228004, 60061228006).
Software: HEAsoft 6.35 (HEASARC 2025), SKIRT \citep{2023A&A...674A.123V},
XSPEC \citep{1996ASPC..101...17A}.


\bibliography{sample701}{}
\bibliographystyle{aasjournalv7}



\end{document}